\documentclass[]{aastex}

\usepackage{emulateapj5}
\usepackage{onecolfloat}
\usepackage{graphicx} 
\usepackage{fancyheadings} 
\usepackage{ulem}
\usepackage{rotating}
\usepackage{lscape}

\newcommand{\ew}{W$_\lambda$}
\newcommand{\EW}{W$_\lambda$}

\newcommand{\msol}{M_\odot}

\newcommand{\Lya}{Ly$\alpha$}
\newcommand{\lya}{Ly$\alpha$}
\newcommand{\lyb}{Ly$\beta$}
\newcommand{\lyg}{Ly$\gamma$}

\newcommand{\lye}{Ly$\epsilon$}

\newcommand{\kms}{km~s$^{-1}$ }
\newcommand{\cm}[1]{\, {\rm cm^{#1}}}
\newcommand{\N}[1]{{N({\rm #1})}}

\newcommand{\sci}[1]{{\rm \; \times \; 10^{#1}}}

\newcommand{\mkms}{{\rm \; km\;s^{-1}}}
\newcommand{\oii}{{\rm O\,II}}
\newcommand{\oiii}{{\rm O\,III}}
\newcommand{\oiv}{{\rm O\,IV}}
\newcommand{\ov}{{\rm O\,V}}
\newcommand{\ovi}{{\rm O\,VI}}
\newcommand{\cii}{{\rm C\,II}}
\newcommand{\ciii}{{\rm C\,III}}
\newcommand{\civ}{{\rm C\,IV}}
\newcommand{\nii}{{\rm N\,II}}
\newcommand{\niii}{{\rm N\,III}}
\newcommand{\niv}{{\rm N\,IV}}
\newcommand{\nv}{{\rm N\,V}}
\newcommand{\siiv}{{\rm Si\,IV}}
\newcommand{\siii}{{\rm Si\,II}}
\newcommand{\siiii}{{\rm Si\,III}}

\newcommand{\spvi}{{\rm S\,VI}}
\newcommand{\neviii}{{\rm Ne\,VIII}}

\begin{document}

\twocolumn[%
\submitted{Accepted to the Astrophysical Journal: August 12, 2004}
\title{Probing the IGM/Galaxy Connection Toward PKS0405-123 I: 
UV Spectroscopy and Metal-Line Systems}

\author{Jason X. Prochaska\altaffilmark{1},
Hsiao-Wen Chen\altaffilmark{2,3},
J. Christopher Howk\altaffilmark{4},
Benjamin J. Weiner\altaffilmark{1},
and John Mulchaey\altaffilmark{5}
}

\begin{abstract}

We present results from an analysis of
{\it Far Ultraviolet Spectroscopic Explorer} 
(FUSE) spectroscopy of the $z_{em}=0.57$
quasar PKS0405--123.  We focus on the intervening metal-line
systems identified along the sightline and investigate their
ionization mechanism, ionization state, and chemical abundances.
Including HST/STIS spectroscopy,
we survey the entire sightline and identify six \ovi\ 
absorbers to a $3 \sigma$ 
equivalent width (EW) limit of 60m\AA. This implies an
incidence $dN/dz = 16^{+9}_{-6}$ consistent with previous \ovi\
studies. In half of the \ovi\ systems we report positive 
detections of \ciii\ suggesting the gas is predominantly photoionized,
has multiple ionization phases, or is in a non-equilibrium state.
This contrasts with the general description of the
warm-hot intergalactic medium (WHIM) as described by numerical simulations
where the gas is predominantly in collisional ionization equilibrium.  
An appreciable fraction of \ovi\ absorbers may therefore have a 
different origin.
We have also searched the sightline for the \neviii\ doublet over
the redshift range $0.2 < z < z_{em}$ which
is a better probe of the WHIM gas at $T > 10^6$\,K.  We find
no positive detections to an EW limit of 80m\AA\ giving
$dN/dz < 40$ at 95$\%$ c.l.

The photoionized metal-line systems exhibit a correlation
between the ionization parameter ($U = \Phi/c n_H$
with $\Phi$ the flux of hydrogen ionizing photons)
and H\,I column density for
$\N{HI} = 10^{14} - 10^{16} \cm{-2}$.
Both the slope and normalization of this correlation match the 
prediction inferred from the results
of Dav\'e and Tripp for the low $z$ \lya\ forest.
In turn, the data show a tentative, unexpected result:
five out of the six photoionized metal-line systems
show a total hydrogen column density within a 
factor of 2 of $10^{18.7} \cm{-2}$.
Finally, the median metallicity $\lbrack$M/H$\rbrack$ of 
twelve $z \sim 0.3$ absorbers
with $\N{HI} > 10^{14} \cm{-2}$ is $\lbrack$M/H$\rbrack$~$> -1.5$ with 
large scatter. This significantly
exceeds the median metallicity of \civ\ and \ovi\ systems at $z \sim 3$.
and therefore requires enrichment of the intergalactic medium
over the past $\approx 10$\,Gyr.



\end{abstract}
]

\altaffiltext{1}{UCO/Lick Observatory; University of California, Santa Cruz;
Santa Cruz, CA 95064; xavier@ucolick.org}
\altaffiltext{2}{Center for Space Research, Massachusetts Institute of 
Technology, Cambridge, MA 02139-4307; hchen@mit.edu}
\altaffiltext{3}{Hubble Fellow}
\altaffiltext{4}{Department of Physics, and Center for Astrophysics and 
Space Sciences, University of California, San Diego, C--0424, La Jolla, 
CA 92093-0424}
\altaffiltext{5}{Observatories of the Carnegie Institution of
Washington, 213 Santa Barbara St., Pasadena, CA 91101}

\pagestyle{fancyplain}
\lhead[\fancyplain{}{\thepage}]{\fancyplain{}{PROCHASKA ET AL.}}
\rhead[\fancyplain{}{Probing the IGM/Galaxy Connection Toward PKS0405-123 I.}]{\fancyplain{}{\thepage}}
\setlength{\headrulewidth=0pt}
\cfoot{}

\section{INTRODUCTION}
\label{sec-intro}

Quasar absorption line studies examine the gas within galaxies
and the intergalactic medium (IGM) between them. 
These studies include analyze of the interstellar medium of young galaxies
at $z>2$ (e.g.\ the damped \lya\ systems) as well as the lines that 
comprise the \lya\ forest.
One of the most notable successes of cold dark matter cosmologies
at $z \sim 3$ is the general
agreement between observations of the \lya\ forest
and cosmological simulations both in terms of the column density
and Doppler width distributions \citep[e.g.][]{miralda96,zhang97}. 
Meanwhile, analyses of metal-line systems describe the enrichment
history of the universe \citep{pettini97,pro03}, test processes of
nucleosynthesis \citep[e.g.][]{lu96,phw03}, and examine the role and
characteristics of winds \citep{aguirre01} and Population III enrichment
\citep{gnedin97,wass00}.

The `fluctuating Gunn-Peterson' paradigm inferred from numerical simulations
describes the majority of the observed IGM as large-scale overdensities with
little correspondence to individual galaxies 
\citep[e.g][]{gnedin98}.  
Although this model reasonably reproduces the number density and gross 
kinematics of the IGM, 
it has not been extensively tested by observations.
This is especially true at low redshift where there are very
few sightlines with high resolution, high quality spectra.
For this reason, alternative models \citep[e.g.\ absorption by gas in galactic
halos or low surface brightness galaxies;][]{linder00,chen01,manning03}
are generally as successful
at explaining the low $z$ IGM 
as the CDM paradigm \citep{dave01}.


We are pursuing an observing
program aimed at addressing key questions related to the low $z$ IGM.  
We have chosen to study the IGM along the sightlines to $z>0.1$
AGN with relatively high signal-to-noise (S/N) far-UV FUSE spectroscopy
and, in many cases, HST/STIS echelle observations.
This UV spectroscopy is complemented by large field of view 
($>20'$ diameter) galaxy surveys compiled with the WFCCD spectrograph
at Las Campanas Observatory.
The principal focus of our program is to identify the
location and physical characteristics 
of low redshift \ovi\ absorbers, whose hot gas content is believed to 
be a significant reservoir of baryons 
\citep{cen99,tripp00,dave01b,fang01}.  
The properties in which we are interested include 
the ionization state, metallicity, and
absorber/galaxy cross-correlation function for \ovi\ gas.
Studies of the \ovi\ absorbers
currently provide the most efficient means of studying the relatively
hot $T \approx 10^{5-7}$K, low density gas termed the
warm-hot intergalactic medium (WHIM).
The WHIM is predicted to contain a large fraction of the baryons at low
$z$ \citep{dave01b}, but its existence is difficult to verify due to its
low density.
In addition to studies on the origin of the WHIM,
the combined datasets
of high quality UV spectroscopy and deep, large field-of-view galaxy 
surveys will examine several aspects of the IGM/galaxy connection.
These issues are intimately related to our understanding of gas and 
galaxies in the local universe.  Through comparisons with hydrodynamic
numerical simulations, we will test and constrain the CDM paradigm for
the IGM and examine processes of galaxy feedback and chemical enrichment.

This paper focuses on the metal-line absorbers identified
in the ultraviolet spectroscopy of PKS0405--123. 
We have focused first on the PKS0405--123 sightline for the following reasons:
(i) It exhibits one of the brightest UV fluxes for a $z>0.5$ quasar.  This 
allowed the STIS science team to obtain high quality UV spectroscopy
from $\lambda \approx 1200-1700$\AA\ and our team to acquire modest
S/N FUV spectroscopy from $\lambda \approx 900-1170$\AA\ with the FUSE
Observatory.
(ii) It exhibits a partial Lyman limit system (i.e.\ an absorber with
$\tau \lesssim 1$ at $\lambda_{rest} = 912$\AA) at $z=0.167$ \citep{chen00}.
In this paper we report on our analysis of the UV spectroscopy and examine
the metal-line systems identified along this sightline. 
and 
(iii) We have composed a 
$\approx 1\square^\circ$ galaxy survey reaching $R \lesssim 20$ with
$\approx 500$ galaxies at $z < z_{em}$.  
Concurrent and forthcoming papers on this sightline will 
(a) report on the physical nature of the absorber 
identified at $z=0.4951$ \citep{howk04}, 
(b) describe the galaxy survey and the connection between galaxies and the
metal-line systems discussed here \citep{pro04}; and 
(c) present an analysis 
of the \lya/galaxy cross-correlation function \citep{chen04}.
Future work will analyze the IGM and the galaxies associated with
it for $\approx 10$ low redshift fields.

The paper is organized as follows:  in $\S$~2 we present the FUSE observations
and a line list of significant absorbers identified along the sightline;
in $\S$~3 we describe our approach to ionization modeling; 
we analyze the ionization state, physical conditions and elemental
abundances of the metal-line systems in $\S$~4; 
$\S$~5 lists the absorbers along the sightline 
with $\N{HI} > 10^{14} \cm{-2}$ which do not display metal transitions;
we discuss the implications of our results for the WHIM, chemical enrichment,
and the general IGM in $\S$~6; and $\S$~7 gives a brief summary.

\begin{table}[ht]\footnotesize
\begin{center}
\caption{{\sc UV SPECTROSCOPY\label{tab:obs}}}
\begin{tabular}{rcccc}
\tableline
\tableline
Instr. &Mode & Wavelength & Exp. & S/N$^a$ \\
\tableline
FUSE     & LWRS  & 900--1180\AA  & 71ks & 5-15 \\
HST/STIS & E140M & 1150--1700\AA & 27ks & 7 \\
\tableline
\end{tabular}
\end{center}
\tablenotetext{a}{Signal-to-noise per resolution element for {\it FUSE} data and
per pixel for HST/STIS.}
\end{table}

\section{OBSERVATIONS, DATA REDUCTION, AND EW ANALYSIS}
\label{sec:redux}

PKS0405--123 was observed for 71ks during Cycle~2 of the FUSE
mission (Program B087; PI: Prochaska).  This program was principally
motivated by the identification of a strong metal-line system at
$z=0.167$ in the HST+STIS/E140M data of PKS0405--123 obtained by the 
STIS science team \citep{williger04}.
In \cite{chen00}, we analyzed this absorption system and argued that it is a
partial Lyman limit system with roughly solar metallicity.  
We then obtained far-UV spectroscopy with the FUSE observatory to
further constrain the physical characteristics of this absorber.
Table~\ref{tab:obs} summarizes the UV spectroscopy acquired with FUSE
and HST/STIS of PKS0405--123 and estimates the signal-to-noise (S/N)
per resolution element.

\begin{table}[ht]\footnotesize
\begin{center}
\caption{{\sc ATOMIC DATA \label{tab:fosc}}}
\begin{tabular}{lcccc}
\tableline
\tableline
Transition &$\lambda$ &$f$ & Ref \\
\tableline
   HI 972  &  972.5368 & 0.029      &  1  \\
   OI 988a &  988.5778 & 0.0005146  &  1  \\
   OI 988b &  988.6549 & 0.007712   &  1  \\
   OI 988  &  988.7734 & 0.04318    &  1  \\
 NIII 989  &  989.7990 & 0.1066     &  1  \\
 SiII 989  &  989.8731 & 0.133      &  1  \\
\tableline
\tablerefs{Key to References -- 1:
\cite{morton91}; 2: \cite{howk00}; 3:
\cite{morton04}; 4: \cite{tripp96}; 5:
\cite{fedchak99}; 
7: \cite{fedchak00}; 9: \cite{schect98}; 11:
\cite{bergs93b}; 12: \cite{wiese01}; 13:
\cite{bergs93}; 14: \cite{bergs94}; 15:
\cite{verner94}; 17: \cite{bergs96}}
\tablecomments{[The complete version of this table is in the electronic edition of
the Journal.  The printed edition contains only a sample.]}
\end{tabular}
\end{center}
\end{table}

\begin{table}[ht]\footnotesize
\begin{center}
\caption{{\sc ADOPTED SOLAR ABUNDANCES\label{tab:solabd}}}
\begin{tabular}{rrr}
\tableline
\tableline
Elm &$\epsilon(X)^a$ & Z \\
\tableline
H & 12.00 &  1\\
C &  8.59 &  6\\
N &  7.93 &  7\\
O &  8.74 &  8\\
Ne &  8.08 & 10\\
Mg &  7.58 & 12\\
Al &  6.49 & 13\\
Si &  7.56 & 14\\
S &  7.20 & 16\\
Fe &  7.50 & 26\\
\tableline
\end{tabular}
\end{center}
\tablenotetext{a}{$\epsilon(X)$ is the logarithmic number density of a
given element scaled such that $\epsilon({\rm H}) = 12.$}
\end{table}

\begin{table*}\footnotesize
\begin{center}
\caption{{\sc EW SUMMARY \label{tab:ewsumm}}}
\begin{tabular}{lrrrcccccc}
\tableline
\tableline
Ion & $\lambda_{obs}$ &$\lambda_{rest}$ & $z_{abs}$ & $W_1$
& $\sigma(W_1)$& $W_2$ & $\sigma(W_2)$ &
$W_f$ & $\sigma(W_f)$ \\
 & (\AA) & (\AA) && (m\AA) & (m\AA) & (m\AA) & (m\AA) & (m\AA) & (m\AA) \\
\tableline
FUSE \\
\tableline
OIV 787& 931.521& 787.711& 0.18257&$   7$&$ 15$&$  42$&$ 12$&$  28$&  9\\
OIV 787& 931.797& 787.711& 0.18292&$ 124$&$ 17$&$ 131$&$ 14$&$ 128$& 11\\
MgX 624& 934.389& 624.950& 0.49514&$  -2$&$ 13$&$ -12$&$ 10$&$  -8$&  8\\
OV 629& 941.457& 629.730& 0.49502&$ 183$&$ 12$&$ 145$&$ 12$&$ 164$&  9\\
SIV 809& 944.893& 809.668& 0.16701&$  13$&$ 16$&$  31$&$ 15$&$  23$& 11\\
OII 834& 973.896& 834.466& 0.16709&$   8$&$ 17$&$  17$&$ 12$&$  14$& 10\\
OIII 832& 985.062& 832.927& 0.18265&$  34$&$ 15$&$   5$&$ 11$&$  15$&  9\\
\tableline
\end{tabular}
\end{center}
\tablecomments{Note that the list is incomplete for wavelengths $< 1000$\AA\ where the data has poor S/N and significant line blending.}
\tablecomments{Columns 4,5 (6,7) refer to the SiC1B (SiC2A) channel for 910\AA$ < \lambda < 990$\AA, LiF1A (LiF2B) for 995\AA$ < \lambda < 1057$\AA, LiF1A for 1057\AA$ < \lambda < 1082$\AA,
 LiF2A for 1082\AA$ < \lambda < 1100$\AA, LiF1B (LiF2A) for 1100\AA$ < \lambda < 1180$\AA, and STIS/E140M for $\lambda > 1190$\AA.} 
\tablecomments{[The complete version of this table is in the electronic edition of the Journal.  The printed edition contains only a sample.]}
\end{table*}

PKS0405--123 was observed with the FUSE observatory \citep{moos00,sahnow00}
using the LWRS apertures with the detectors in
TTAG mode during one visit beginning UT 05 October 2001.
We used {\it ttag\_combine} to concatenate the photon event lists
into a single file and then
processed this file with the CalFUSE pipeline (v2.4).  The
reduction proceeded without any significant warnings and the standard
series of output files was created.
We binned these data arrays by three `pixels' ($\approx 0.028$\AA\ per bin)
and then traced the continuum
of PKS0405--123 in each channel separately by fitting 
Legendre polynomials to regions of unabsorbed quasar continuum.
These normalized spectra form the dataset for our analysis. 
The atomic data considered in this paper is tabulated in Table~\ref{tab:fosc}
and Table~\ref{tab:solabd} lists our assumed solar abundance data 
compiled by \cite{grvss99}
and revised by \cite{holweger01} for C,N, and O.

Redward of 1000\AA, we identified all $3\sigma$ features in the LiF1A and
LiF2A channels.  There is moderate absorption from Galactic molecular hydrogen
along this sightline; we attribute $\approx 50\%$ of the features
at $\lambda = 1000-1150$\AA\ to H$_2$ or Galactic metal-line transitions.
Table~\ref{tab:ewsumm} lists the observed wavelength, transition name, 
absorption redshift, rest equivalent width $W_\lambda$, and the statistical 
error in $W_\lambda$ from each channel where
the transition is observed at reasonable S/N.
Columns 4,5 (6,7) refer to the SiC1B (SiC2A) channel for 
910\AA$ < \lambda < 990$\AA, LiF1A (LiF2B) for 995\AA$ < \lambda < 1057$\AA, 
LiF1A for 1057\AA$ < \lambda < 1082$\AA, 
LiF2A for 1082\AA$ < \lambda < 1100$\AA, 
LiF1B (LiF2A) for 1100\AA$ < \lambda < 1180$\AA, and STIS/E140M for
$\lambda > 1190$\AA\ (see Chen \& Prochaska 2000 for details of the 
STIS dataset and reduction).  
Note that there are offsets
between the wavelength solutions for each FUSE channel.  We corrected for these
offsets with a cross-correlation analysis allowing for a zero point 
and a first order term.
The latter term, while small, is required to match the spectrum
over the entire spectral range of each channel.
This procedure only corrects for relative errors;  a comparison
of H\,I Lyman lines and metal lines from the STIS and FUSE spectra indicates
the absolute wavelength calibration is accurate to better than 10\,\kms.
The final columns in Table~\ref{tab:ewsumm} list
the adopted $W_\lambda$ value and uncertainty corresponding to 
the variance weighted mean for transitions with multiple measurements.  
In general, the values derived from multiple channels
show good agreement.
For several transitions (e.g.\ Galactic P\,II 1152), however, the individual
$W_\lambda$ values differ by $>3\sigma$ statistical significance.  
In most cases, this may be attributed to fixed pattern noise in the 
detectors or continuum uncertainties.  The transitions showing
these discrepant cases were
either ignored or we adopted an uncertainty
equal to the spread in the two values. 

\vskip 1.0in

\begin{figure}[ht]
\begin{center}
\includegraphics[height=3.6in,angle=90]{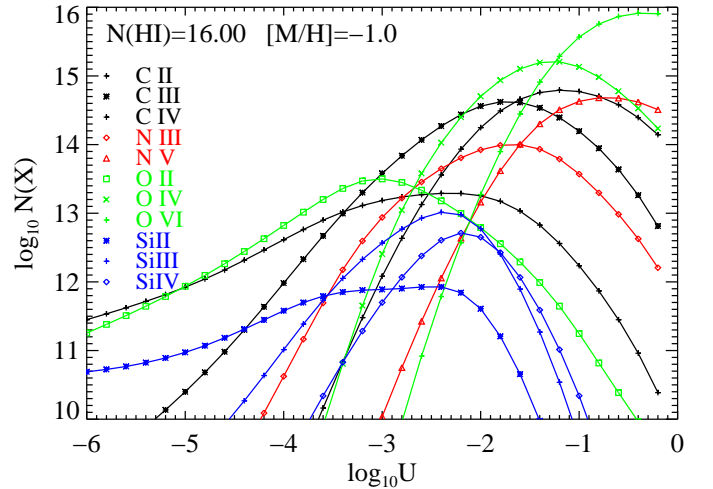}
\caption{Predicted ionic column densities for a photoionized
gas as a function of ionization parameter $U$.  The curves are the results
of a series of calculations with the Cloudy software package \citep{ferland01}
assuming (i) a quasar-only extragalactic UV background (model~Q) at $z=0$;
(ii) a total H\,I column density $\N{HI} = 10^{16} \cm{-2}$;
and (iii) solar relative abundances scaled to a metallicity of 1/10 solar.
}
\label{fig:cldyex}
\end{center}
\end{figure}

\section{IONIZATION MODELING}
\label{sec:ionmod}

\subsection{Photoionization}

In this paper, we examine gas likely to be significantly
ionized.  For those systems which show multiple ionic species, it is possible
to constrain the ionization mechanism 
(i.e.\ photoionization vs.\ collisional ionization) 
and ionization state of the gas.  In general, we expect the gas to be significantly
photoionized, either by 
an extragalactic UV background (EUVB) radiation field
or a local radiation source 
(e.g.\ proximity to a starbursting galaxy).  For the shape of the EUVB field, 
we consider two models generated by \cite{haardt04} using
their CUBA software package:
quasar-only (Q) and quasars+galaxies (QG).
These models input the luminosity function of AGN and UV bright galaxies
for a given redshift
and then calculate the attenuation of the mean intensity by the intergalactic
medium.  The majority of previous studies have adopted
model~Q for the EUVB field.  In our analysis, 
we will consider this model as the default spectrum, and we note
that ongoing programs studying the UV emission of low $z$ galaxies
(e.g.\ GALEX) will provide a more accurate assessment of the proper
EUVB model.  Where relevant, we describe the effects of including 
a softer radiation field than model~Q.

In practice we calculate solutions for a plane-parallel slab of gas using
the Cloudy software package \citep{ferland01}.  
Because all of the absorbers along the PKS0405--123 sightline are optically thin to 
the radiation field, the calculation is simplified and the models are fully
described by 
(i) the ionization parameter $U \equiv \Phi/cn_H$, with $\Phi$ the 
flux of hydrogen ionizing photons and $n_H$ the volume density of hydrogen;
(ii) the shape of the ionizing radiation field; 
and (iii) the metallicity of the gas [M/H].
The models are most sensitive to variations in the ionization
parameter while the shape of the
radiation field and the gas metallicity play more modest roles.
Throughout the analysis we assume constant density (typically an arbitrary
$\log n_H = -1$) and constrain the plane-parallel slab to have the
thickness required to yield the observed $\N{HI}$ value given the specific
ionization state of the gas.

We have constructed
grids of these photoionization models varying $U$, [M/H], and the EUVB 
model for a range of redshifts and $\N{HI}$ values.  
An example of one set of models is shown in Figure~\ref{fig:cldyex}
where we have assumed EUVB model Q at $z=0$ and $\N{HI} = 10^{16} \cm{-2}$.  
The lines trace predicted
ionic column densities vs.\ $\log U$ for a series of ions 
in a gas with 1/10 solar 
metallicity (i.e.\ [M/H]~$= -1$) and solar relative abundances
(Table~\ref{tab:solabd}).  Low-ion species (e.g.\ Si\,II) show less
variation at $\log U<-3$ because they trace the H\,I
gas to first order and the models assume a fixed $\N{HI}$ value.
In contrast, high-ion species show large values only for higher
ionization parameters.  

The general effect of adopting a softer radiation field (e.g.\ EUVB
model~QG) is that one requires a higher ionization parameter to
attain significant column densities of \oiv,\ov, etc.
In turn, higher $U$ values imply lower neutral fractions and therefore
larger $\N{H}$ values for a fixed (observed) $\N{HI}$
value which -- contrary to our initial expectation -- 
predicts lower metallicities
($\approx 0.3 - 0.5$~dex lower) for absorbers dominated by high-ions.
Systems characterized by a mixture of low
and intermediate ions, however, show minimal differences in their
elemental abundances as a function of the EUVB model.

Our general approach for a given absorber is to (1) constrain $U$ 
with one or more ionic ratios, ideally of the same element
(e.g.\ $\N{\siii}/\N{\siiii}$); (2) determine if a self-consistent
solution is allowed with all of the ions included; and
(3) estimate the ionization parameter, 
metallicity, relative abundances, and
physical conditions (e.g.\ density, temperature) of the gas from the best
ionization model.
This technique is dominated by systematic error, e.g.\ the
limited constraints on the EUVB model, the simplifying
assumption of constant density, etc.
With current model sophistication and typical observational constraints, we
do not believe one can achieve better than 0.2--0.3~dex precision 
on elemental abundances or 0.2~dex for relative abundances. 
Therefore, we take a cautious approach when
comparing the observations against the photoionization models
and adopt conservative errors.

\begin{figure}[ht]
\begin{center}
\includegraphics[height=3.6in,angle=90]{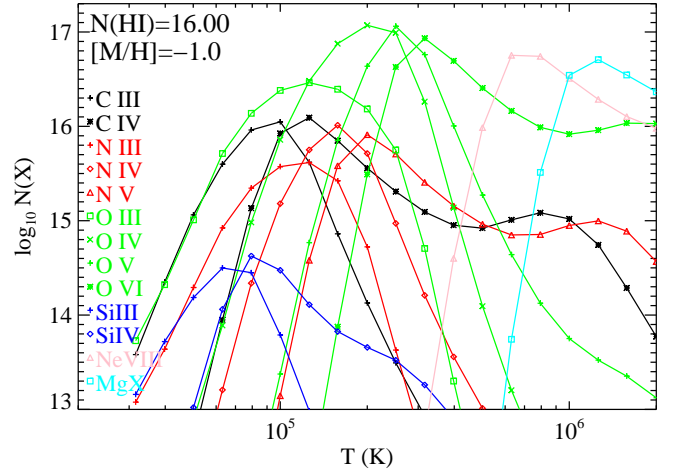}
\caption{
Predicted ionic column densities for a collisionally ionized gas
in equilibrium as a function of gas temperature $T$.  The curves were
calculated with the Cloudy software package 
assuming (i) no ionizing radiation field;
(ii) $\N{HI} = 10^{16} \cm{-2}$;
and (iii) solar relative abundances scaled to a metallicity of 1/10 solar.
}
\label{fig:coll}
\end{center}
\end{figure}

\subsection{Collisional Ionization}

For gas at relatively high temperatures $(T > 5 \sci{4}$\,K), collisional
ionization can be the dominant ionization process for a number of high-ions
observed in quasar absorption line 
systems (e.g.\ \civ, \nv, \ovi). 
In Figure~\ref{fig:coll} we plot abundance curves vs.\ temperature 
under the assumption of collisional ionization equilibrium 
\citep[CIE;][]{sutherland93} as calculated by the Cloudy package.
We have assumed 1/10 solar metallicity and plot the predicted column 
densities for several ions assuming $\N{HI} = 10^{16} \cm{-2}$ and solar
relative abundances (Table~\ref{tab:solabd}).
Collisional ionization is important for higher ionization states
of Si and C 
at $T \approx 5\sci{4} - 2\sci{5}$~K whereas O and N become
highly ionized at $T \approx 10^5 - 10^6$~K.  There are several
key differences between the curves presented in Figures~\ref{fig:cldyex}
and \ref{fig:coll} which direct our discussion on the ionization mechanism
of the metal-line systems toward PKS0405--123.  In particular, note that the
\ciii\ and \ovi\ ions do not coexist with significant values
in the collisional ionization
solutions yet both ions show large column densities for $\log U \approx -1.5$
in the photoionization model.  A similar difference is noted for \niii. 
In addition, the relative abundances of the individual O ions show modest
differences in the two models, e.g.\ photoionization tends to predict
smaller differences between $\N{\oiv}$ and $\N{\ovi}$ than collisional
ionization.
Another key difference between the two scenarios is temperature
of the gas: CIE models generally predict much higher temperature. 
Therefore, the Doppler width ($b$ value) of the absorption 
can provide a discriminant
between the two scenarios, especially when the neutral hydrogen gas
shows $b < 30 \mkms$.

\subsection{Multi-phase Scenarios and Other Ionization States}

If one is limited to the observation of only a handful of ions, however, it
may be difficult to distinguish between a gas with an ionization phase
dictated by photoionization versus collisional ionization.  For example,
a CIE model with $T \approx 2\sci{5}$~K predicts similar
relative ionic column densities of \ovi\ and \nv\ 
as a photoionization model with $\log U > -1$.
In these cases, degenerate solutions exist which limit our constraints
on the physical properties of the gas.  Furthermore, the ionization
mechanisms are not exclusive.  
For the absorbers considered in this paper, one expects that a radiation
field is always present and a pure CIE model is somewhat
unrealistic (although the effects of photoionization may be minor).  
In the following section, our approach will be to assess single-phase
ionization models first and only consider multi-phase scenarios as
required by the observations.

There are also examples in the literature of ionized gas which is
not well explained by equilibrium models of collisional or
photionization.  The best-known examples of non-equilibrium gas
involve gas in the interstellar medium of galaxies at temperatures
where the cooling rate is extremely high, including $T~10^5$ K gas
studied through lines from the Li-like ions \ion{O}{6}, \ion{N}{5},
and \ion{C}{4}.  The prevailing conclusion is that this gas is not in
equilibrium and arises as a result of conductive heating, radiative
shocks, and/or turbulent mixing layers (see Spitzer 1996 for a
summary).  Models of these phenomena predict a wide range of
gas properties and diagnostics dependent on the
assumed initial conditions and timescales for cooling, heating, and
mixing.   In this paper, we will restrict our analysis equilibrium
models of collisional and photoionization.  In part, this is because
we have very few measurements of C\,IV and Si\,IV for these absorbers.
More importantly, examples of conduction heating and radiative shocks
are more likely to occur in dense, star forming regions;  such
conditions are unlikely for the absorption systems considered here.
Finally, we will find that the systems are well described
by a single or two-phase equilibrium model.  Nevertheless, 
it is important to stress
that our results are sensitive to the presumption of equilibrium
modeling.

\section{ANALYSIS OF THE METAL-LINE SYSTEMS}
\label{sec:analysis}

This section presents an analysis of all of the metal-line systems identified 
along the sightline to PKS0405--123.  
Because of the complicated line-spread functions of the spectrographs,
we derive the $\N{HI}$ values and Doppler parameters for the H\,I
gas using a curve-of-growth (COG) analysis.
For the metal-line transitions, we report rest-frame equivalent
widths, column densities
calculated using the apparent optical 
depth method \citep[AODM;][]{sav91,jenkins96},
and statistical errors for $3 \sigma$ significant transitions.
At FUSE resolution, the majority of weak transitions
are free of significant line saturation and the AODM approach
should give accurate column densities.  The technique 
has the added advantage that 
the values are free of the parametric modeling inherent to line-profile
fitting.   In general, upper limits reflect 
$2\sigma$ statistical limits and lower limits
signify probable line saturation.
We identify each system by a redshift which 
corresponds to the centroid of the hydrogen Lyman series or 
the peak optical depth of the metal-line transitions.
Throughout this section, the data plotted are from the 
SiC2A, LiF1A, and LiF2A channels or the STIS instrument.
For the velocity plot profiles, we indicate blends with coincident
absorption lines (primarily Galactic H$_2$) as dotted orange lines.

We caution that this section is quite detailed.  The casual reader
may wish to skip to the summary table presented at the end of the
section.

\begin{figure}[ht]
\begin{center}
\includegraphics[width=3.6in]{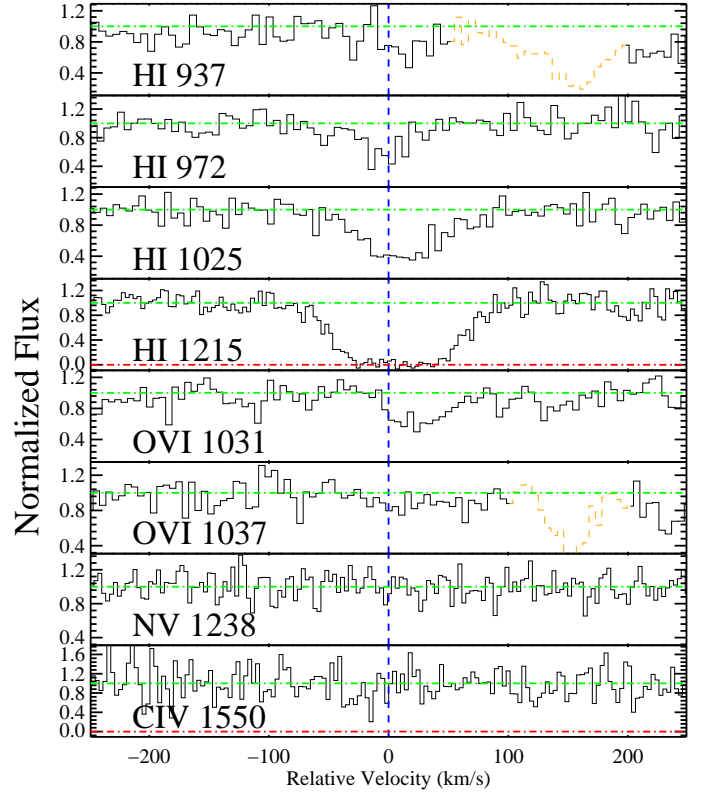}
\caption{
Velocity profiles of the Lyman series and metal-line transitions
analyzed for the absorption system at $z=0.09180$.
In this and the following velocity profile plots, the dotted orange
lines denote blends from coincident transitions and the dash-dot red
line indicates the zero level.
}
\label{fig:z0918}
\end{center}
\end{figure}

\begin{table}[ht]\footnotesize
\begin{center}
\caption{{\sc IONIC COLUMN DENSITIES FOR THE ABSORBER AT $z$=0.09180\label{tab:z0.0918}}}
\begin{tabular}{lcccc}
\tableline
\tableline
Ion &$\lambda$ (\AA) & EW (m\AA) & AODM & $N_{adopt}$ \\
\tableline
HI & & & & $14.52 \pm 0.05$\\
CIV&1550.7700&$<  49 $&$ < 13.70$ &$< 13.70$ \\
NV&1238.8210&$<  20 $&$ < 13.15$ &$< 13.15$ \\
OVI&1031.9261&$  73 \pm   8 $&$ 13.82 \pm 0.05 $&$ 13.83 \pm 0.04$ \\
OVI&1037.6167&$  38 \pm  10 $&$ 13.83 \pm 0.11 $&$ $ \\
\tableline
\end{tabular}
\end{center}
\end{table}
 
\begin{figure}[ht]
\begin{center}
\includegraphics[height=3.6in,angle=90]{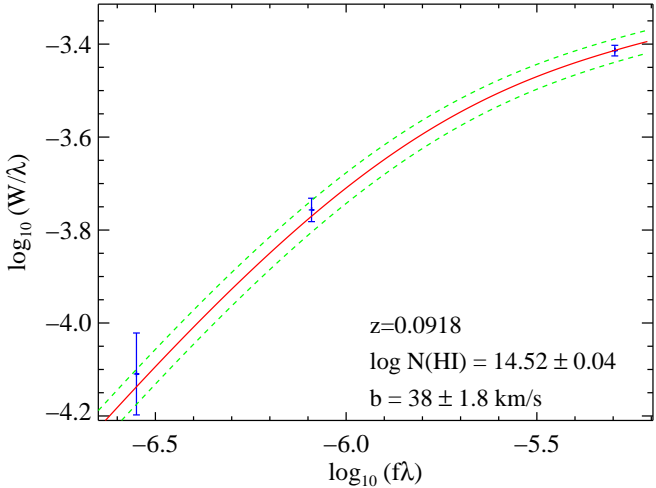}
\caption{
Curve-of-growth analysis of the 
Lyman series for the gas at $z=0.09180$.
}
\label{fig:0918COG}
\end{center}
\end{figure}

\subsection{$z=0.09180$}

The lowest redshift metal-line system identified toward
PKS0405--123 is an \ovi\ system at $z=0.0918$.
Both transitions of the O\,VI doublet are detected and give a column density
$\N{\ovi} = 10^{13.8 \pm 0.04} \cm{-2}$ (Table~\ref{tab:z0.0918}).
The STIS+FUSE spectra reveal H\,I transitions for
\Lya, \lyb, \lyg, and a marginal \lye\ detection (Figure~\ref{fig:z0918}).
The results of a COG analysis are presented in Figure~\ref{fig:0918COG}. 
Minimizing $\chi^2$ for the \EW values as a function of $f\lambda$, we derive
$\N{HI} = 10^{14.52 \pm 0.04} \cm{-2}$ and $b = 38 \pm 2 \mkms$.

Unfortunately, the C\,III 977 transition for this absorber is blended with
the Lyman series of the metal-line system
at $z=0.167$.  As noted above ($\S$~\ref{sec:ionmod}), \ciii\ is
particularly valuable for evaluating the 
ionization mechanism of O\,VI absorbers.
With only a limited set of ionic column densities for this absorber, 
both CIE and
photoionization solutions are allowed although the non-detection of \nv\
places tight constraints on the temperature and ionization
parameter.   For collisional ionization equilibrium, the upper limit to 
$\N{\nv}/\N{\ovi}$ sets a lower limit to the temperature of
$T > 2.5 \sci{5}$K assuming [N/O]~$>-0.5$\,dex.  
This temperature is roughly consistent with the
Doppler parameter measured for the H\,I gas ($b \approx 40 \mkms$).
We also note that the O\,VI profile is offset by $\approx +20 \mkms$
from the centroid of the H\,I profile.  Perhaps the \ovi\
gas is related to a more tenuous H\,I component.
At the least, the kinematics raise the likelihood that this
is a multi-phase absorber.

For photoionization (EUVB model~Q), the $\N{\nv}/\N{\ovi}$ limit restricts
the ionization parameter  $\log U > {\rm [N/O]} - 1.1$.  
Assuming [N/O]~$= 0$ and $J_{912} = 2\sci{-23}$ 
this implies a gas density $n_H < 3\sci10^{-5} \cm{-3}$ which is consistent
with the volume density 
for absorbers with $\N{HI} \approx 10^{14.5} \cm{-2}$ 
predicted by numerical simulations \citep[e.g.][]{dave01b}. 
In addition, this implies an absorber size $\ell > 30$\,kpc.

We can place a lower limit to the O/H metallicity by considering the
minimum \ovi/H\,I ionization correction.  This gives
[O/H]~$>-1.4$ for photoionization and [O/H]~$>-2.2$ for collisional
ionization.  Note that these are strict lower limits because we have assumed
all of the observed H\,I gas is associated with the O\,VI absorber.

\begin{figure}[ht]
\begin{center}
\includegraphics[width=3.6in]{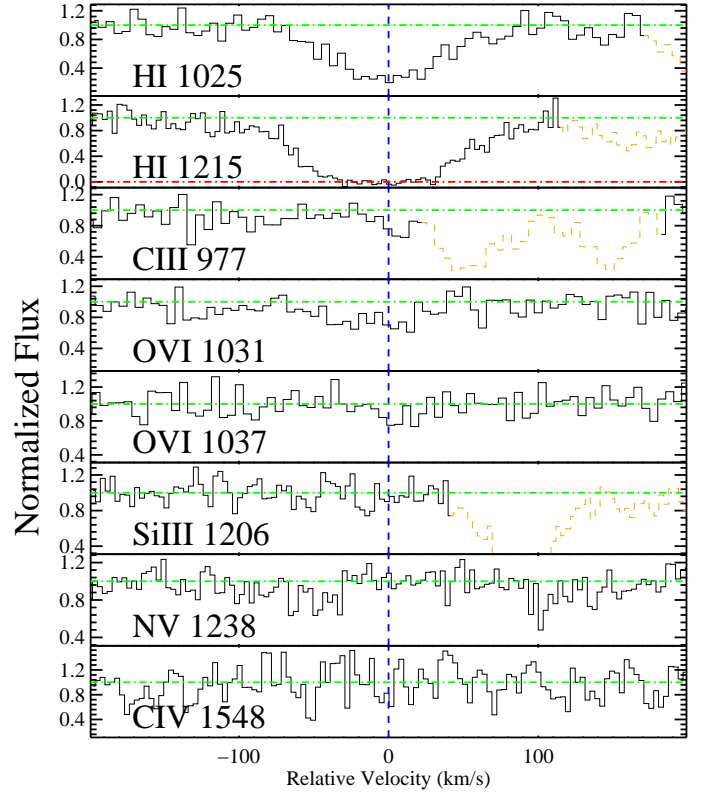}
\caption{
Velocity profiles of the Lyman series and metal-line transitions
analyzed for the absorption system at $z=0.09658$.
}
\label{fig:0965}
\end{center}
\end{figure}

\begin{table}[ht]\footnotesize
\begin{center}
\caption{{\sc IONIC COLUMN DENSITIES FOR THE ABSORBER AT $z$=0.09658\label{tab:z0.0966}}}
\begin{tabular}{lcccc}
\tableline
\tableline
Ion &$\lambda$ (\AA) & EW (m\AA) & AODM & $N_{adopt}$ \\
\tableline
HI & & & & $14.65 \pm 0.05$\\
CIII& 977.0200&$  26 \pm   7 $&$ 12.68 \pm 0.12 $&$ 12.68 \pm 0.12$ \\
CIV&1548.1950&$<  53 $&$ < 13.30$ &$< 13.30$ \\
NV&1238.8210&$<  16 $&$ < 13.06$ &$< 13.06$ \\
OVI&1031.9261&$  71 \pm   9 $&$ 13.81 \pm 0.06 $&$ 13.7 \pm 0.2$ \\
OVI&1037.6167&$<  17 $&$ < 13.63$ &$$ \\
SiIII&1206.5000&$<  14 $&$ < 12.02$ &$< 12.02$ \\
\tableline
\end{tabular}
\end{center}
\end{table}

\subsection{$z=0.09658$}

At $z=0.09658$ we identify an absorber with similar characteristics 
to the \ovi\ system at $z=0.09180$.  We measure a nearly 
identical \ovi\ column
density from the \ovi~1031 profile but note the \ovi~1037 transition
is not detected at the $3\sigma$ level (Figure~\ref{fig:0965}).  
We also caution that continuum placement is especially important for this
absorber.
Finally, one notes that the \ovi~1031 profile extends to $v \approx -50\mkms$
whereas \ovi~1037 is confined to $v > -20 \mkms$ (in both the LiF2A and LiF1B
channels).
In any event, we adopt a final column 
density $\N{\ovi} = 10^{13.7 \pm 0.2} \cm{-2}$
based on the column densities of both profiles.
For now, we include this absorber in our sample and note that FUSE Cycle~4
observations of PKS0405--123 (PI: Howk) should resolve these concerns.
A curve-of-growth analysis shows this
system has nearly the same $\N{HI}$ value as the $z=0.09180$ absorber
and a comparable Doppler width $b = 40 \pm 2 \mkms$.

Although the C\,III~977 transition is blended with a 
coincident transition, the absorption at $v \approx +10\mkms$
is most likely related to this absorber and we report a value for
its column density by integrating from $v \approx -10$ to $+20\mkms$.  
If the \ciii\
gas arises in the same ionization phase as the \ovi\ gas, 
the absorber cannot be explained by a single-phase 
CIE model\footnote{If the absorption at $v < -20 \mkms$ in the \ovi~1031
profile is confirmed in both \ovi\ transitions (the current S/N is 
insufficient to make a definitive statement), then one may need to consider
a multi-phase or non-equilibrium model to explain the kinematic differences
between the \ciii\ and \ovi\ ions.}.
The collisional ionization models (Figure~\ref{fig:coll}) 
do yield a solution with 
log~$\N{\ciii}/\N{\ovi} \approx -1$ at $T \approx 2\sci{5}$~K, yet 
the model also predicts \\
log~$\N{\civ}/\N{\ciii} > 1$ in contradiction with the observations.  Similarly,
this model would predict a \nv\ detection unless [N/O]~$< -0.8$~dex.
In contrast, a photoionization model with $\log U \approx -1.2$ 
matches the observed $\N{\ciii}/\N{\ovi}$ ratio and also 
predicts \civ\ and \nv\ column
densities consistent with the observed upper limits (e.g. Figure~\ref{fig:cldyex}).
Therefore, we contend
photoionization is the principal ionization mechanism for this system.
Adopting $\log U = -1.2 \pm 0.2$, we estimate the gas metallicity 
[O/H]~$=-1.5 \pm 0.3$~dex. 

\begin{figure}[ht]
\begin{center}
\includegraphics[width=3.6in]{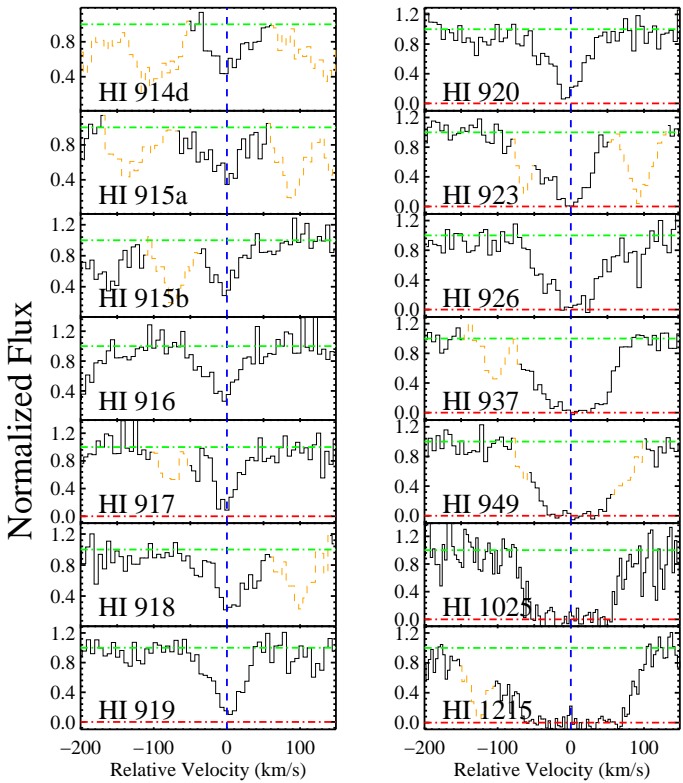}
\caption{
Velocity profiles of the Lyman series for the partial Lyman limit
at $z=0.16710$ toward PKS0405--123.
}
\label{fig:1671HI}
\end{center}
\end{figure}

\begin{figure}[ht]
\begin{center}
\includegraphics[height=3.8in,angle=90]{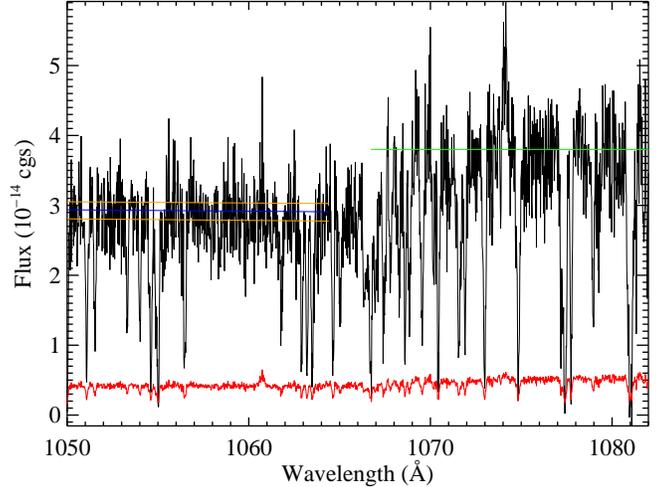}
\caption{
Lyman limit of the absorption system at $z=0.16710$.   The green line
depicts our estimate of the flux redward of the Lyman limit and the
blue line indicates our estimate of the flux blueward of the limit
with a conservative error estimate (yellow lines).  This flux decrement corresponds
to $\log \N{HI} = 16.45 \pm 0.05$.
}
\label{fig:1671LL}
\end{center}
\end{figure}

\begin{figure}[ht]
\begin{center}
\includegraphics[height=3.6in,angle=90]{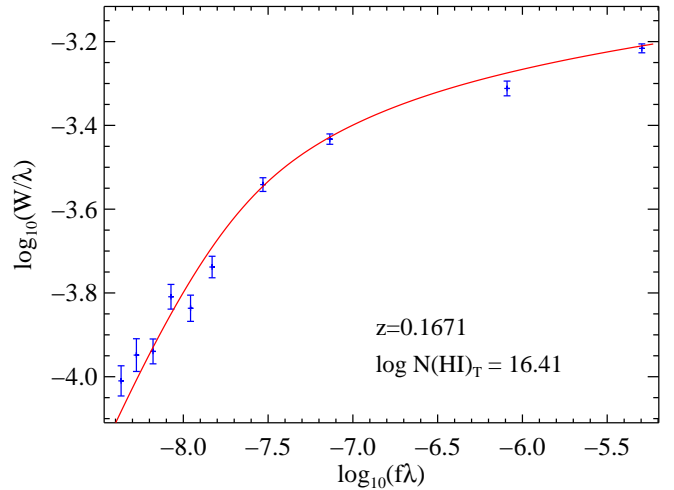}
\caption{
Curve-of-growth analysis of the 
Lyman series for the gas at $z=0.16710$.  In this analysis, we have
assumed two components with $\log \N{HI} = 16.35$ and 15.5
and with $40 \mkms$ separation.
}
\label{fig:1671COG}
\end{center}
\end{figure}

\subsection{$z=0.16710$}

This absorption system was the principal motivation for the 
pursuit of FUSE observations of PKS0405--123.  In particular, these data
allow an analysis of the Lyman series (Figure~\ref{fig:1671HI}) and
therefore a determination of its $\N{HI}$ value.
The absorber is a partial Lyman limit system, i.e.,
the optical depth at $\lambda_{rest} = 912$\AA\ is $\tau_{912} \lesssim 1$.  
Therefore, one
can independently solve for the total $\N{HI}$ value of this absorber through
a COG analysis and a measurement of $\tau_{912}$.  The $\N{HI}$ value
obtained from $\tau_{912}$ is not sensitive to the Doppler parameter, 
instrument line-spread function, or any component structure within the absorber.
In short, the $\tau_{912}$ measurement tightly constrains
the total HI column density with precision
limited only by uncertainty in the quasar continuum. 
In Figure~\ref{fig:1671LL}, we present the FUSE data covering the
Lyman limit of the $z=0.1671$ system.  The overplotted lines mark our 
assessment of the quasar
continuum redward and blueward of $\lambda = 1066$\AA.  Even with a very 
conservative estimate of the continuum error, the uncertainty
in $\N{HI}$ is $<10\%$ and we find $\N{HI} = 10^{16.45 \pm 0.05} \cm{-2}$.

Performing a COG analysis, we find that a single component model is
a poor match to the observed \ew\ values.  
Examining the low-ion profiles, we note that the \cii\ and \siii\ 
profiles show two components at $v_1 \approx +5 \mkms$ and $v_2 \approx -35 \mkms$
in Figure~\ref{fig:1671LL}.  This suggests the H\,I profile is a combination
of two components and that a two-component COG analysis is warranted.
Indeed, a two component solution with 
$N_1({\rm H\,I}) = 10^{16.35} \cm{-2}$ and $N_2({\rm H\,I}) = 10^{15.5} \cm{-2}$ 
and component separation $\delta v= 40 \mkms$ is a good description of the EW
observations (Figure~\ref{fig:1671COG}).
In the following analysis, we will not treat the components separately
because (i) the $N_2({\rm H\,I})$ value is not well constrained and
(ii) the components show similar relative ionic ratios (with the possible
exception of \nii).

\begin{figure}
\plotone{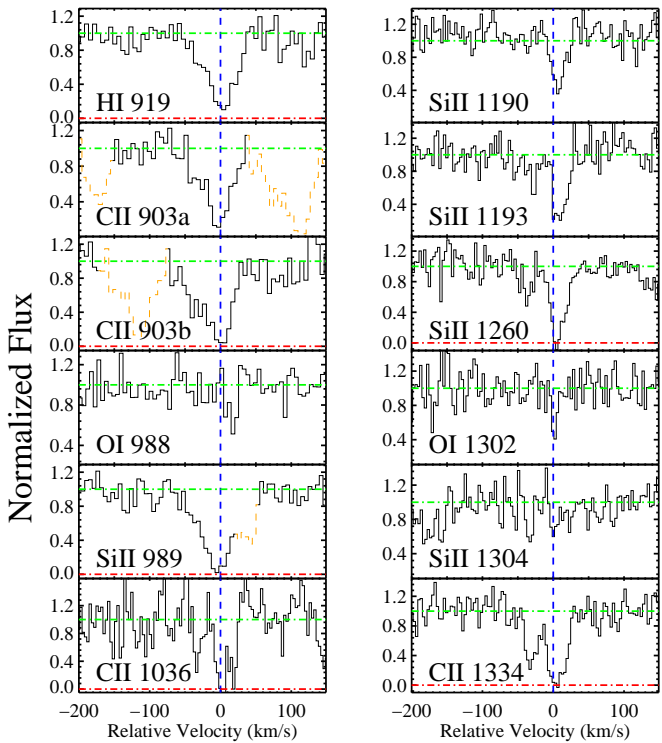}
\caption{
Velocity profiles of the low-ion metal transitions for the partial Lyman limit
at $z=0.16710$ toward PKS0405--123.
}
\label{fig:1671low}
\end{figure}

\begin{figure}
\plotone{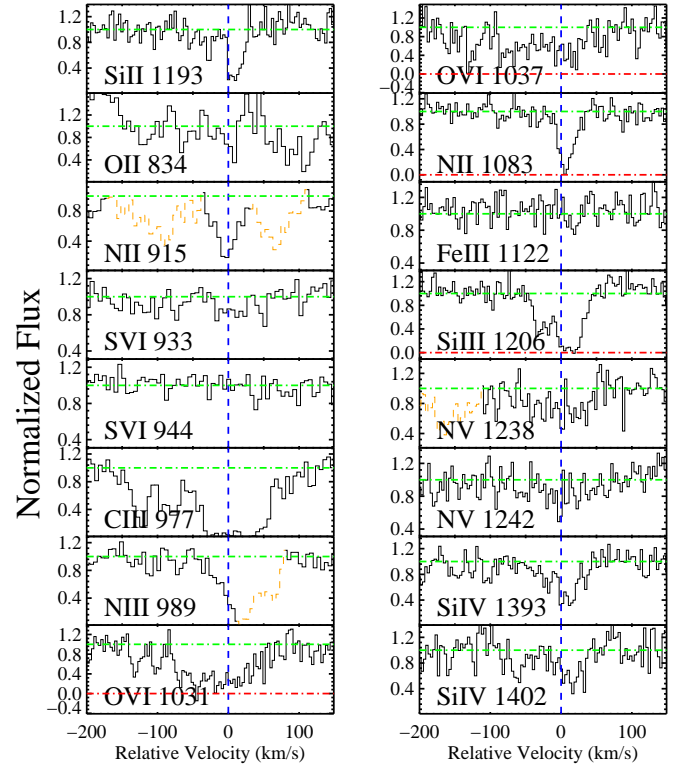}
\caption{
Velocity profiles of the intermediate and high-ion 
metal transitions for the partial Lyman limit
at $z=0.16710$ toward PKS0405--123.
}
\label{fig:1671high}
\end{figure}

The metal-line profiles for the low and high-ion species from the FUSE
and STIS datasets are presented in Figures~\ref{fig:1671low} 
and \ref{fig:1671high}.
The FUSE observations contribute a number of new metal transitions, although
several of the most valuable profiles (e.g.\ O\,II~832, O\,III~834, N\,III~989)
are unfortunately blended with strong Galactic transitions.
The column density measurements and limits for all of the observed transitions
are presented in Table~\ref{tab:z0.1671}.  
One notes that we report a $3\sigma$ detection for the O\,I~988 transition
and a $>2.5\sigma$ detection for the 
O\,I~1302 transition with consistent column densities.  At present, we
consider these tentative detections to be spurious primarily because the implied
O metallicity is super-solar even prior to ionization corrections. 
Furthermore, the O\,I~1302 profile is more narrow than the other low-ion profiles
and is offset from the O\,I~988 profile\footnote{The latter effect could be a small
calibration error}.
We hope that additional, planned FUSE observations of PKS0405--123 (PI: Howk)
will confirm (or contradict) this tentative O\,I measurement.  
For now, we adopt an upper limit to 
$\N{O^0}$ of $10^{13.85} \cm{-2}$.

\begin{table}[ht]\footnotesize
\begin{center}
\caption{{\sc IONIC COLUMN DENSITIES FOR THE ABSORBER AT $z$=0.16710\label{tab:z0.1671}}}
\begin{tabular}{lcccc}
\tableline
\tableline
Ion &$\lambda$ (\AA) & EW (m\AA) & AODM & $N_{adopt}$ \\
\tableline
HI & & & & $16.45 \pm 0.07$\\
CII& 903.6240&$ 125 \pm   8 $&$ > 14.24$ &$> 14.35$ \\
CII& 903.9620&$ 158 \pm   8 $&$ > 14.13$ &$$ \\
CII&1334.5323&$ 197 \pm  14 $&$ > 14.35$ &$$ \\
CIII& 977.0200&$ 374 \pm   8 $&$ > 14.27$ &$> 14.27$ \\
NII& 915.6130&$ 104 \pm   7 $&$ > 14.17$ &$> 14.24$ \\
NII&1083.9900&$  93 \pm   9 $&$ > 14.24$ &$$ \\
NIII& 989.7990&$<  15 $&$ < 14.59$ &$< 14.59$ \\
NV&1238.8210&$ 114 \pm  20 $&$ 13.84 \pm 0.07 $&$ 13.89 \pm 0.05$ \\
NV&1242.8040&$  78 \pm  13 $&$ 13.95 \pm 0.07 $&$ $ \\
OI& 988.7730&$  20 \pm   6 $&$ 13.84 \pm 0.11 $&$ < 13.85$ \\
OI&1302.1685&$  30 \pm  10 $&$ 13.74 \pm 0.14 $&$ $ \\
OII& 834.4655&$<  20 $&$ < 13.65$ &$< 13.65$ \\
OVI&1031.9261&$ 377 \pm  20 $&$ > 14.74$ &$14.78 \pm 0.07$ \\
OVI&1037.6167&$ 240 \pm  23 $&$ 14.78 \pm 0.07 $&$ $ \\
SiII&1190.4158&$  26 \pm   7 $&$ 13.15 \pm 0.09 $&$ 13.30 \pm 0.04$ \\
SiII&1193.2897&$ 105 \pm  11 $&$ 13.42 \pm 0.05 $&$ $ \\
SiII&1260.4221&$ 140 \pm  11 $&$ > 13.36$ &$$ \\
SiII&1304.3702&$  19 \pm   6 $&$ 13.23 \pm 0.14 $&$ $ \\
SiIII&1206.5000&$ 204 \pm  10 $&$ > 13.39$ &$> 13.39$ \\
SiIV&1393.7550&$ 140 \pm  14 $&$ 13.33 \pm 0.04 $&$ 13.34 \pm 0.04$ \\
SiIV&1402.7700&$  92 \pm  16 $&$ 13.44 \pm 0.08 $&$ $ \\
SIII&1190.2080&$<  13 $&$ < 13.83$ &$< 13.83$ \\
SIV& 809.6680&$<  22 $&$ < 13.80$ &$< 13.80$ \\
SIV&1062.6620&$<  18 $&$ < 13.89$ &$$ \\
SVI& 933.3780&$  24 \pm   8 $&$ 12.91 \pm 0.13 $&$ 12.91 \pm 0.13$ \\
SVI& 944.5230&$<  11 $&$ < 12.95$ &$$ \\
FeII&1144.9379&$<   9 $&$ < 13.15$ &$< 13.15$ \\
FeIII&1122.5260&$<  16 $&$ < 13.60$ &$< 13.60$ \\
\tableline
\end{tabular}
\end{center}
\end{table}

\cite{chen00} noted that no equilibrium photoionization or collisional 
ionization model
could match the full set of ionic column densities measured for 
this absorber.  In
particular, the ionic column densities of the \ovi\ and \nv\ 
ions are too large to reconcile with the low and 
intermediate ions in a single ionization phase
(e.g.\ consider Figures~\ref{fig:cldyex},\ref{fig:coll}).  
This conclusion is supported
by the differences between velocity centroids and 
Doppler widths measured for the N\,V and O\,VI profiles and those of
the other metal-line profiles.  Therefore,
\cite{chen00} analyzed the low and intermediate ions independent of 
the \ovi\ and \nv\
gas and we follow this approach. 
Specifically, we examine the ionization state required to give the
$\N{\ovi}/\N{\nv}$ ratio and independently the physical conditions
that yield the lower ions.

Consider first the high-ions under the premise of collisional ionization. 
To reproduce the $\N{\ovi}/\N{\nv}$ ratio with
solar relative abundances, this implies $T \approx 2.5 \sci{5}$\,K.
At this temperature, this CIE phase would have a relatively small ionic
column density for all other ions except \civ\ and \spvi.
The CIE model does predict $\log \N{\spvi} = 12.9$
which matches the column density derived from S\,VI~$\lambda 933$, 
although this measurement has a large uncertainty.
Future observations of C\,IV~$\lambda\lambda 1548, 1550$ 
and S\,VI would verify the
assumption of CIE and tightly constrain the temperature of this phase.
In passing, we note that the metallicity of the gas assuming
$T=2.5 \sci{5}$\,K and $\N{HI} = 10^{16} \cm{-2}$ is [O/H]~$\approx -1$\,dex.

\begin{figure}
\includegraphics[height=3.6in,angle=90]{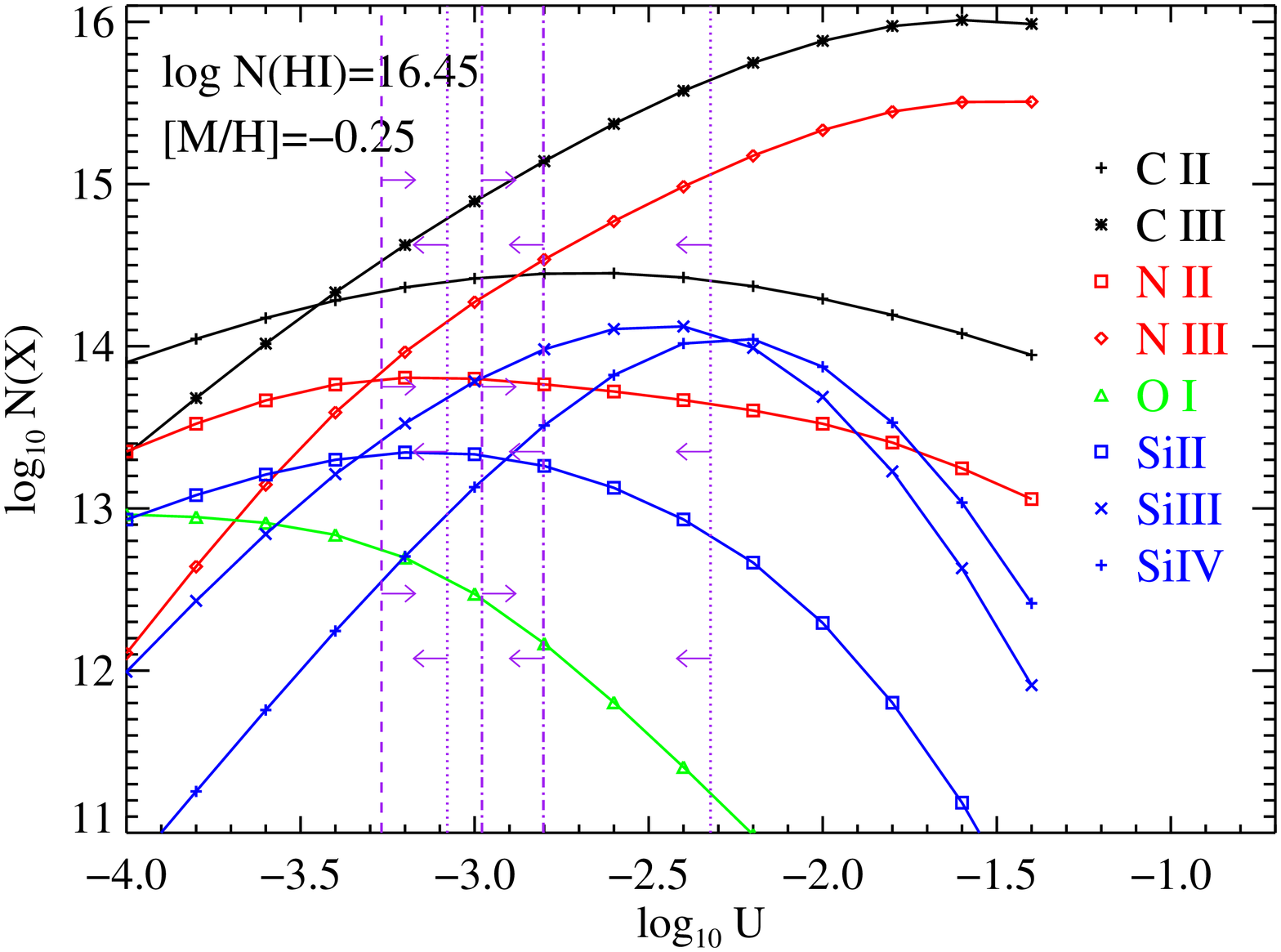}
\caption{
Predicted ionic column densities for the absorber at $z=0.16710$
assuming photoionization and 
a quasar-only extragalactic UV background (model~Q) at $z=0.15$,
a total H\,I column density $\N{HI} = 10^{16.45} \cm{-2}$,
and solar relative abundances scaled to a metallicity [M/H]~$=-0.25$\,dex.
The purple dashed line shows a lower limit to $U$ based on the 
$\N{\siii}/\N{\siiii}$ upper limit whereas the dotted purple lines show
upper limits to $U$ based on limits to the
$\N{\nii}/\N{\niii}$ and $\N{\siii}/\N{\siiv}$ ratios.
Finally the $\N{\siii}/\N{\siiv}$ value and $1\sigma$ error
constrain $U$ to the region indicated by the dash-dot purple lines.
}
\label{fig:1671Cldy}
\end{figure}

We now focus on the lower ionization phase which we expect is 
predominantly photoionized.
Figure~\ref{fig:1671Cldy} presents the predicted ionic column densities
for a range of ions assuming a plane-parallel slab of gas with
$\N{HI} = 10^{16.5} \cm{-2}$, metallicity [M/H]~$= -0.25$~dex, and the
EUVB model Q at $z=0.2$ (see $\S$~\ref{sec:ionmod}).
Comparisons of the model predictions with the 
observed ionic ratios constrain the ionization parameter
and provide estimates to the ionization corrections. 
The tightest constraints on the ionization parameter comes from
the Si and N ions.
The $\N{\siii}/\N{\siiii},$ $\N{\nii}/\N{\niii},$ $ \N{\siii}/\N{\siiv}$ limits and
the $\N{\siii}/\N{\siiv}$ measurement imply
the dashed (lower limits), dotted (upper limits),
and dash-dot (measurements) lines overplotted on Figure~\ref{fig:1671Cldy}.
Formally, no single ionization parameter satisfies all of the constraints.
Given that the Cloudy calculations are based on simplified assumptions
(e.g.\ constant density, single temperature),  the results are in good
agreement\footnote{It is worth noting that adopting a softer ionizing spectrum 
(e.g.\ Model~QG) would worsen the disagreement between the N and Si ions.}
with $\log U = -2.9 \pm 0.2$\,dex.
Applying the appropriate ionization corrections,
we find [Si/H]~$\approx -0.25 \pm 0.2$~dex based on the $\N{\siii}/\N{HI}$ ratio
which is relatively insensitive to $U$ at these values. 
We adopt a conservative
error on [Si/H] reflecting the systematic uncertainties of
photoionization modeling.  
One may formally refer to this metallicity as a lower limit because
an unknown fraction of H\,I gas must be attributed to the more highly ionized
gas phase.  We suspect, however, that this would imply
at most a 0.3~dex increase in [Si/H].

\begin{table}[ht]\footnotesize
\begin{center}
\caption{{\sc ELEMENTAL ABUNDANCES FOR THE ABSORBER AT $z$=0.16710\label{tab:z0.1671x}}}
\begin{tabular}{lccc}
\tableline
\tableline
Ion & [X/H] & [X/Si$^+$] \\
\tableline
C$^{+}$ & $>-0.32$ & $>-0.03$ \\
C$^{++}$ & $>-0.88$ & $>-0.59$ \\
N$^{+}$ & $> 0.19$ & $> 0.48$ \\
N$^{++}$ & $< 0.07$ & $< 0.35$ \\
N$^{+4}$ & $ 1.86\pm0.30$ & $ 2.15$ \\
O$^{0}$ & $< 1.13$ & $< 1.41$ \\
O$^{+}$ & $<-1.33$ & $<-1.04$ \\
O$^{+5}$ & $ 3.88\pm0.49$ & $ 4.17$ \\
Si$^{+}$ & $-0.29\pm0.02$ & $ 0.00$ \\
Si$^{++}$ & $>-0.65$ & $>-0.36$ \\
Si$^{+3}$ & $-0.04\pm0.20$ & $ 0.25$ \\
S$^{++}$ & $< 0.07$ & $< 0.36$ \\
S$^{+3}$ & $< 0.95$ & $< 1.23$ \\
S$^{+5}$ & $ 1.34\pm0.37$ & $ 1.63$ \\
Fe$^{+}$ & $< 0.78$ & $< 1.06$ \\
Fe$^{++}$ & $<-0.14$ & $< 0.15$ \\
\tableline
\end{tabular}
\end{center}
\end{table}

Table~\ref{tab:z0.1671x} lists the [X/H] and [X/Si$^+$]
values for all of the ions observed in this partial LLS assuming
$U_{best} = 10^{-3.0 \pm 0.1}$.  
For illustration, we include \nv\ and \ovi\ to demonstrate 
these ions cannot
be in photoionization equilibrium with the other ions because one predicts very
low values compared to the observations.
For S and Fe (marked by non-detections), the calculated upper limits on [S/H] and
[Fe/H] are consistent with expectation.  
The C ions are also in reasonable agreement although the lower limits
from \cii\ nearly conflict with the Si abundance under the presumption
of solar relative abundances.  As noted above, if the O\,I transitions are 
true detections they imply a large O abundance and a correspondingly
large O/Si ratio.
If future observations yield an $\N{O^0}$ value near our upper limit,
the super-solar O/Si ratios might indicate gas enriched by massive 
$(M>15 \msol)$ Type~II SN \citep{ww95}.

Because the FUSE+STIS datasets provide full coverage of the dominant 
Si and N ions indicated by the photoionization modeling, 
it is possible to
calculate the N/Si ratio largely independent of ionization corrections
for the photoionized gas\footnote{Note $\N{\niv}$ and $\N{\nv}$ are assumed
to be smaller than the upper limit obtained for $\N{\niii}$}:

\begin{equation}
{\rm \frac{N}{Si}} =  \frac{\N{\nii} + \N{\niii}}{\N{\siii} + \N{\siiii} + 
\N{\siiv}}
\quad .
\end{equation}

\noindent 
Summing over the column densities and assuming reasonable values
for the limits,
we find log(N/Si)~$\gtrsim +0.5$ or [N/Si]~$\gtrsim +0.2$~dex.  
We derive a similar value by considering only the $\N{\nii}/\N{\siii}$ ratio
which has a small dependence on $U$ for $\log U < -2.7$\,dex.
This N/Si overabundance is a surprising result; there are few examples 
of enhanced N abundances in the local universe. 
We also note that the O and C observations are consistent with
enhancements of these elements relative to Si suggesting an 
overabundance of all the light elements.
One of the few sites where
large N abundances are observed are quasars \citep{osmer80,bentz04}.
Given the properties of high metallicity, enhanced light elements, 
and substantial
separation from any bright galaxy \citep{pro04}, 
one may speculate that this gas was related to AGN activity.

\begin{figure}
\plotone{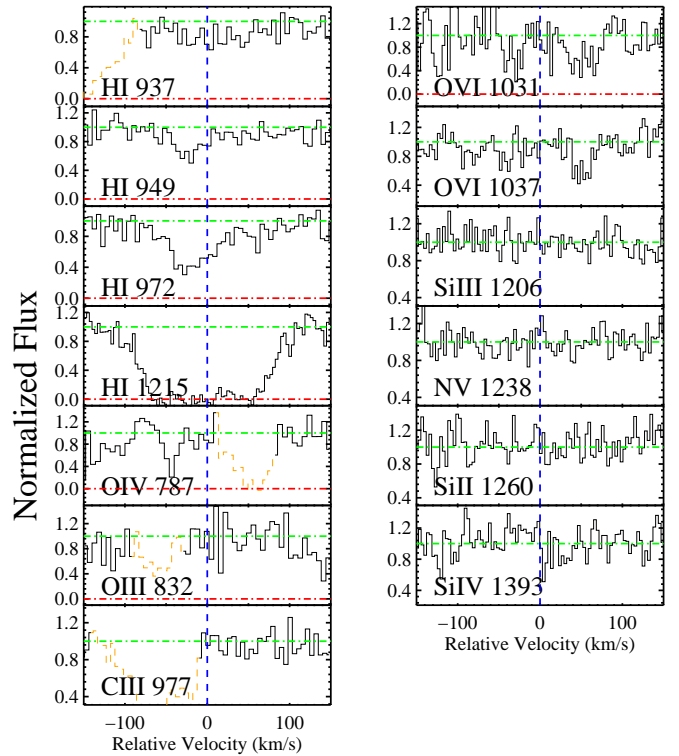}
\caption{
Velocity profiles of the Lyman series and metal-line transitions
analyzed for the pair of absorption systems at $z=0.183$.
The dotted line in the figure corresponds to $z=0.18270$.
}
\label{fig:1827}
\end{figure}

\subsection{$z=0.18250$ and $z=0.18290$}

The system at $z=0.183$ is notable for showing large 
$\N{\ovi}/\N{HI}$
variations in a pair of absorbers separated by 
$\delta v < 100$~\kms.
Figure~\ref{fig:1827} plots the H\,I and metal-line profiles for the
two absorbers with $z=0.18250$ ($v= -25 \mkms$) and $z=0.18290$ ($v= +50 \mkms$).
We performed separate COG analyses 
and derived H\,I column
densities $\N{HI} = 10^{14.50 \pm 0.05} \cm{-2}$ and $10^{14.08 \pm 0.1} \cm{-2}$
and Doppler parameters $b = 27 \pm 1 \mkms$ and $b=26 \pm 5 \mkms$
respectively.
The only metal-line absorption that we confidently identify with this system
is \ovi\ gas related to the $z=0.1829$ absorber.  There is a hint of
\ovi\ absorption at the velocity corresponding to the $z=0.1825$ absorber,
but the measured equivalent width is less than a $3\sigma$ detection.
At this redshift, the FUSE data provide coverage of the 
O\,III~832 and O\,IV~787 transitions of this absorber.
Unfortunately, the O\,IV~787 transition is significantly blended with
an H$_2$ transition.  The O\,III~832 region is also partially
contaminated but provides a meaningful upper limit on the \oiii\ 
column density (Tables~\ref{tab:0.1825}, \ref{tab:0.1829}).

\begin{table}[ht]\footnotesize
\begin{center}
\caption{{\sc IONIC COLUMN DENSITIES FOR THE ABSORBER AT $z$=0.18250\label{tab:0.1825}}}
\begin{tabular}{lcccc}
\tableline
\tableline
Ion &$\lambda$ (\AA) & EW (m\AA) & AODM & $N_{adopt}$ \\
\tableline
HI & & & & $14.90 \pm 0.05$\\
NV&1238.8210&$<  14 $&$ < 12.99$ &$< 12.99$ \\
OII& 834.4655&$<  21 $&$ < 14.55$ &$< 14.55$ \\
OIII& 832.9270&$<  18 $&$ < 13.73$ &$< 13.73$ \\
OIV& 787.7110&$<  19 $&$ < 13.83$ &$< 13.83$ \\
OVI&1031.9261&$<  28 $&$ < 13.77$ &$< 13.77$ \\
OVI&1037.6167&$<  23 $&$ < 13.95$ &$$ \\
SiII&1260.4221&$<  15 $&$ < 12.21$ &$< 12.21$ \\
SiIII&1206.5000&$<  14 $&$ < 11.99$ &$< 11.99$ \\
SiIV&1393.7550&$<  28 $&$ < 12.64$ &$< 12.64$ \\
\tableline
\end{tabular}
\end{center}
\end{table}

\begin{table}[ht]\footnotesize
\begin{center}
\caption{{\sc IONIC COLUMN DENSITIES FOR THE ABSORBER AT $z$=0.18290\label{tab:0.1829}}}
\begin{tabular}{lcccc}
\tableline
\tableline
Ion &$\lambda$ (\AA) & EW (m\AA) & AODM & $N_{adopt}$ \\
\tableline
HI & & & & $14.08 \pm 0.10$\\
CIII& 977.0200&$<  11 $&$ < 12.42$ &$< 12.42$ \\
NV&1238.8210&$<  15 $&$ < 13.05$ &$< 13.05$ \\
OIII& 832.9270&$<  25 $&$ < 13.79$ &$< 13.79$ \\
OIV& 787.7110&$<  22 $&$ < 14.61$ &$< 14.61$ \\
OVI&1031.9261&$  72 \pm  18 $&$ 13.94 \pm 0.12 $&$ 13.96 \pm 0.07$ \\
OVI&1037.6167&$  44 \pm   9 $&$ 13.97 \pm 0.09 $&$ $ \\
SiIII&1206.5000&$<  14 $&$ < 12.01$ &$< 12.01$ \\
SiIV&1393.7550&$<  27 $&$ < 12.68$ &$< 12.68$ \\
\tableline
\end{tabular}
\end{center}
\end{table}

We detect no other ions for these two absorbers.  Most importantly,
we place an upper limit
to $\N{\ciii} < 10^{12.4} \cm{-2}$ for the
absorber at $z=0.18290$
\footnote{The C\,III~977 profile is blended with an H$_2$
transition at the velocity corresponding to $z=0.18250$.}.
If the gas is photoionized, the $\N{\ovi}/\N{\ciii}$ ratio requires an ionization
parameter $\log U > -1$ for solar relative abundances
which implies [O/H]~$>-1$
(e.g.\ Figure~\ref{fig:cldyex}).
Under the assumption of collisional ionization,
however,  the $\N{\ciii}/\N{\ovi}$ limit requires $T > 2 \sci{5}$~K and 
suggests metallicity $-2 < $~[O/H]~$< -1$.  
If we ignore the Doppler parameter of the H\,I gas which is uncertain, 
both solutions are allowed but imply temperatures for the gas
which differs by more than one order of magnitude. 
A detection or sensitive limit to the $\N{\civ}$ value would likely resolve
the degeneracy; this measurement could be achieved with future HST+STIS
observations of PKS0405--123. 

Irrespective of the ionization mechanism,
the absorbers are notable for exhibiting $\N{\ovi}/\N{HI}$ ratios differing
by more than an order of magnitude in gas separated by $\delta v < 100 \mkms$
\citep[see also][]{savage02}.
Under the assumption of photoionization, the variation in $\N{\ovi}/\N{HI}$ 
can only be explained through large differences in [M/H] and/or
$n_H$.  For example, the $z=0.18250$ absorber
may have substantially higher gas density and a correspondingly lower
ionization parameter.
In this case, however, the models predict $\N{\oiv}$ 
values inconsistent with the observed upper limits. Therefore, if the 
gas is photoionized
the absorber at $z=0.18290$ has at least $10\times$ higher metallicity.
This would indicate a remarkable level of inhomogeneity in metal enrichment
and would be particularly surprising given the $z=0.18290$ absorber has 
lower H\,I column density.
If collisional ionization is the dominant mechanism, then the $\N{\ovi}/\N{HI}$
ratios imply large temperature differences if the two absorbers 
have comparable metallicity (e.g.\ $T \gtrsim 10^6$K
for the gas at $z=0.18250$ and $T \approx 3\sci{5}$K for 
the gas at $z=0.18290)$.  If we require $T > 2 \sci{5}$\,K for the
$z=0.18290$ system based on the non-detection of C\,III, then we derive
a temperature for the $z=0.18250$ absorber which is inconsistent at
greater than $5 \sigma$ significance from the COG analysis.  Therefore,
it is likely that a CIE model would also require a large 
($>0.3$\,dex) difference in [O/H].

\begin{figure}
\plotone{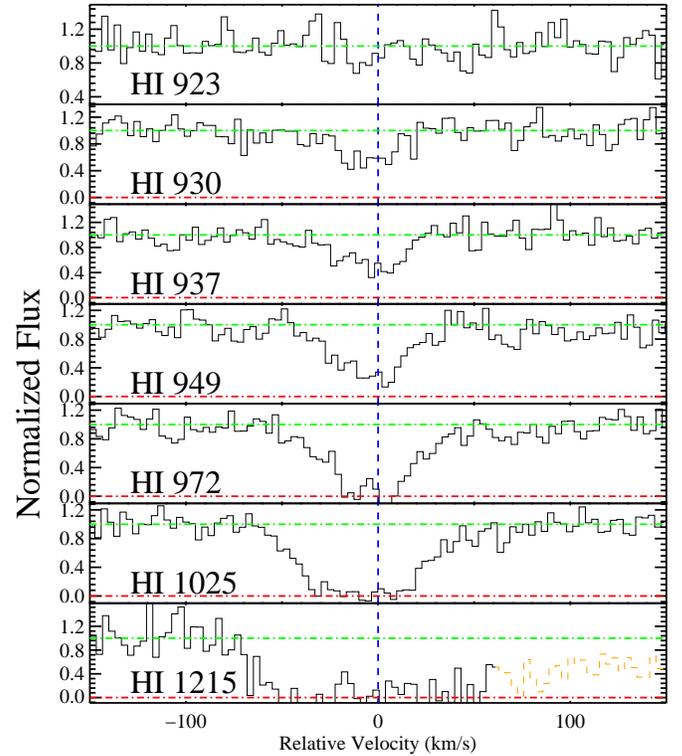}
\caption{
Velocity profiles of the Lyman series for the absorption system
at $z=0.36080$ toward PKS0405--123.
}
\label{fig:3608HI}
\end{figure}

\begin{figure}
\plotone{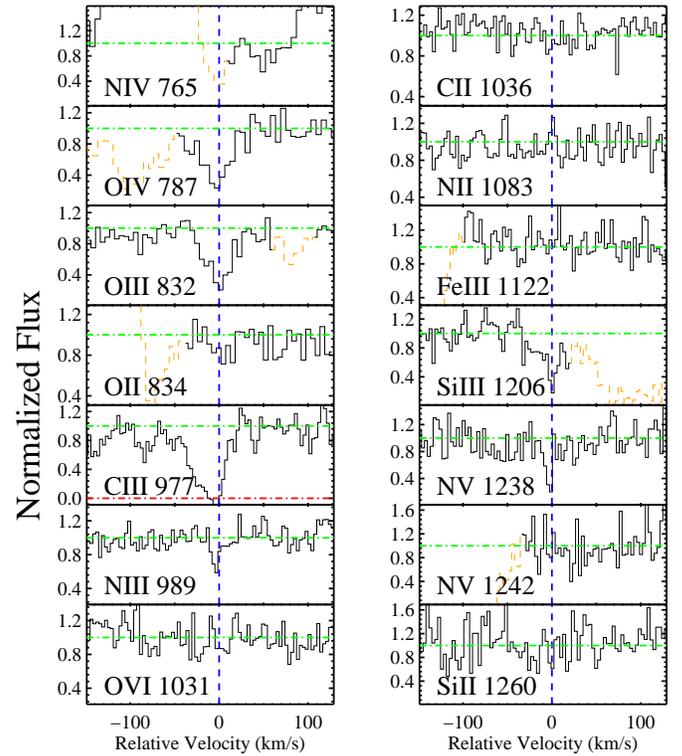}
\caption{
Velocity profiles of the metal-line transitions for the absorption system
at $z=0.36080$.
}
\label{fig:3608mtl}
\end{figure}

\subsection{$z=0.36080$}

There is a complex, highly ionized absorption system at $z=0.36080$
toward PKS0405--123.
The FUSE+STIS datasets show \Lya-Ly6 absorption 
(Figure~\ref{fig:3608HI}) which shows a hint of two closely
separated components.  A single component COG analysis, however, provides
a good description of the data and reveals
$\N{HI} = 10^{15.12 \pm 0.05} \cm{-2}$ and $b = 27 \pm 1 \mkms$.
We identify a series of metal-line transitions including
absorption from \oiii\, \oiv, and \ciii\ (Figure~\ref{fig:3608mtl},
Table~\ref{tab:z0.3608}).  
We also detect N\,IV~765 and Si\,III~1206 but report upper limits because of 
line blending.
Finally, there is a $2\sigma$ detection of the N\,III~989 transition and
a feature at the expected position of N\,V~1238.  We suspect the latter feature
is related to fixed pattern noise or detector artifact 
because the observed optical depth
is only marginally consistent with the null detection of N\,V~1242 and because
of the implied N abundance:  [N/Si]~$> 0.5$\,dex.
Interestingly, this absorber shows no low-ion gas and also no O\,VI absorption.
We predict the absorber will show significant C\,IV absorption 
and, therefore, may be an analog 
to high $z$ C\,IV systems with comparable H\,I column density
\citep[e.g.][]{sargent88}.  Indeed, this is apparently the case as
Williger et al.\ (2004) find $\log \N{CIV} = 13.76 \pm 0.16$ based on
their analysis of STIS G230M archival data.

\begin{table}[ht]\footnotesize
\begin{center}
\caption{{\sc IONIC COLUMN DENSITIES FOR THE ABSORBER AT $z$=0.36080\label{tab:z0.3608}}}
\begin{tabular}{lcccc}
\tableline
\tableline
Ion &$\lambda$ (\AA) & EW (m\AA) & AODM & $N_{adopt}$ \\
\tableline
HI & & & & $15.12 \pm 0.05$\\
CII&1036.3367&$<  13 $&$ < 13.21$ &$< 13.21$ \\
CIII& 977.0200&$ 152 \pm   7 $&$ > 13.78$ &$> 13.78$ \\
NII&1083.9900&$<  19 $&$ < 13.47$ &$< 13.47$ \\
NIII& 989.7990&$<  11 $&$ < 13.31$ &$< 13.31$ \\
NIV& 765.1480&$<   8 $&$ < 13.22$ &$< 13.22$ \\
NV&1238.8210&$<  23 $&$ < 13.52$ &$< 13.52$ \\
NV&1242.8040&$<  32 $&$ < 13.75$ &$$ \\
OII& 834.4655&$<  10 $&$ < 13.25$ &$< 13.25$ \\
OIII& 832.9270&$  72 \pm   5 $&$ > 14.17$ &$> 14.17$ \\
OIV& 787.7110&$  75 \pm   7 $&$ > 14.26$ &$> 14.26$ \\
OVI&1031.9261&$<  16 $&$ < 13.30$ &$< 13.30$ \\
SiII&1260.4221&$<  28 $&$ < 12.55$ &$< 12.55$ \\
SiIII&1206.5000&$<  19 $&$ < 12.64$ &$< 12.64$ \\
FeIII&1122.5260&$<  19 $&$ < 13.67$ &$< 13.67$ \\
\tableline
\end{tabular}
\end{center}
\end{table}

Examining the collisional ionization solutions described 
by Figure~\ref{fig:coll},
a solution with $T \approx 1.3\sci{5}$~K matches the relative column densities
of the O ions. 
We do not favor this scenario, however, for several reasons.  First, collisional
ionization implies $\N{\ciii}/\N{\oiii} \approx -0.8$ yet we observe
$\N{\ciii}/\N{\oiii} > -0.4$ under the assumption 
that the C\,III~977 profile requires
larger corrections for line-saturation than 
O\,III~832.  Second, the temperature is inconsistent
at $> 5\sigma$ with the value implied by the measured H\,I Doppler parameter.
Third, the predicted $\N{\niii}/\N{\niv}$ ratio is only 
marginally consistent with the observed limits.
Although none of these arguments is decisive, together they imply
collisional ionization is not the dominant ionization process.
A conclusive test (e.g.\ observations of the C\,IV doublet)
would be valuable as CIE models imply
significantly lower metallicity and correspondingly
higher $\N{H}$ than photoionization.  

\begin{figure}
\includegraphics[height=3.6in,angle=90]{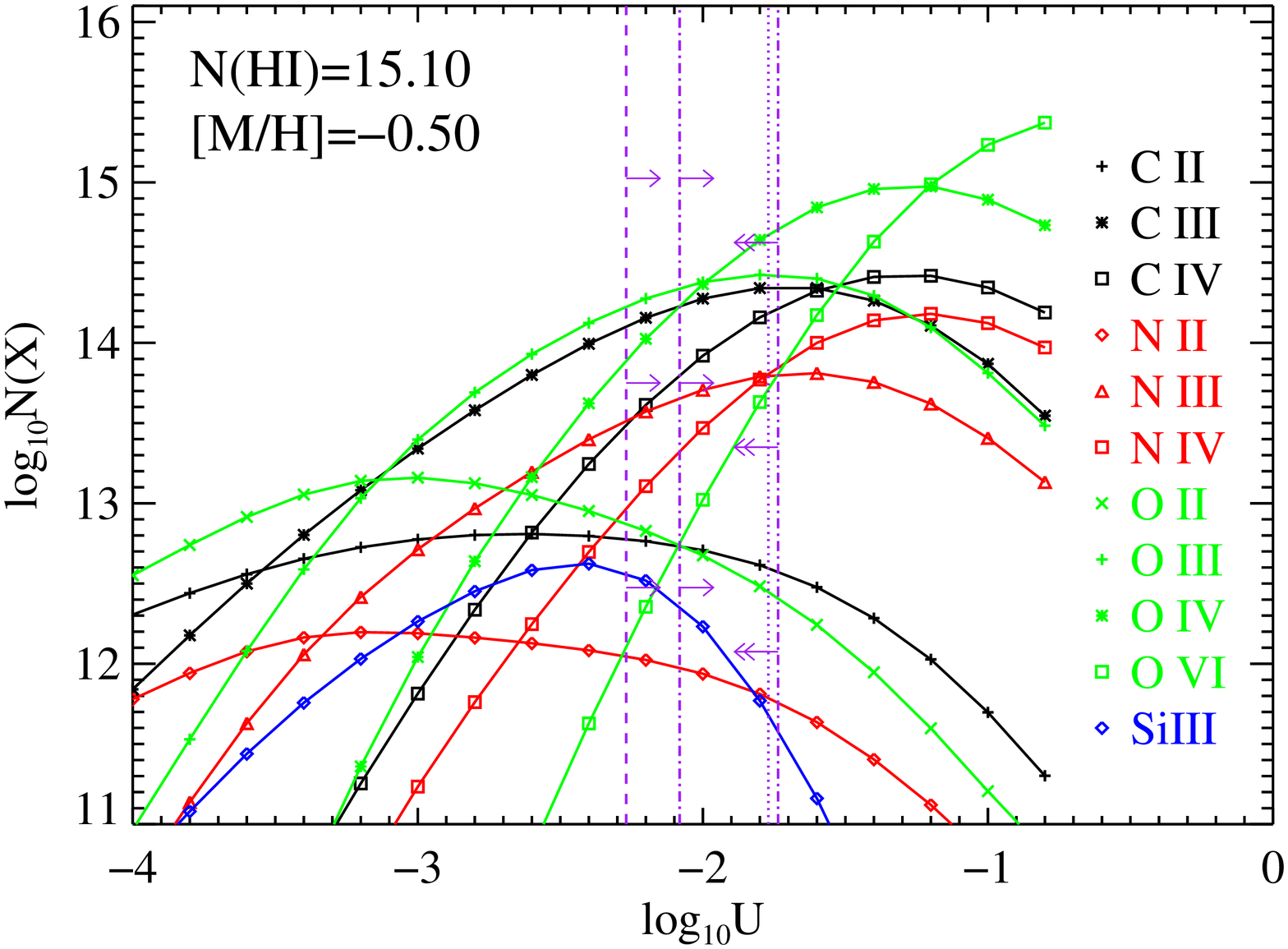}
\caption{
Predicted ionic column densities for the absorber at $z=0.36080$
assuming photoionization and 
a quasar-only extragalactic UV background (model~Q) at $z=0.35$,
a total H\,I column density $\N{HI} = 10^{15.10} \cm{-2}$,
and solar relative abundances scaled to a metallicity [M/H]~$=-0.50$\,dex.
The purple dashed line shows a lower limit to $U$ based on the 
$\N{\oii}/\N{\oiv}$ upper limit whereas the dotted purple line shows
an upper limits to $U$ based on the lower limit to $\N{\oiv}/\N{\ovi}$.
Finally the $\N{\oiii}/\N{\oiv}$ value and $1\sigma$ error
constrain $U$ to the region indicated by the dash-dot purple lines.
}
\label{fig:3608Cldy}
\end{figure}

Assuming photoionization,
the measurements and limits on the column densities of the O ions
tightly constrain the ionization state 
of this gas (Figure~\ref{fig:3608Cldy}).  
Treating the \oiii\ and \oiv\ column densities
as lower limits because of line saturation, the resulting
$\N{\oii}/\N{\oiv}$ and $\N{\oiii}/\N{\ovi}$ 
limits require $-2.1 < \log U < -1.7$.  
If we adopt $\log [ \N{\oiii}/\N{\oiv}] = 0.1 \pm 0.2$,
this places a similar constraint on the $U$ parameter.
In Table~\ref{tab:z0.3608x}, we list the ions analyzed for this absorber
and [X/H] values corresponding to $U_{best} = 10^{-1.9 \pm 0.2}$.
The principal result is that the lower limits to \oiii\ and \oiv\
imply [O/H]~$\gtrsim -0.7$,
an enrichment level consistent with the lower limit
to the \ciii\ column density. 

\begin{table}[ht]\footnotesize
\begin{center}
\caption{{\sc ELEMENTAL ABUNDANCES FOR THE ABSORBER AT $z$=0.36080\label{tab:z0.3608x}}}
\begin{tabular}{lccc}
\tableline
\tableline
Ion & [X/H] & [X/Si$^+$] \\
\tableline
C$^{+}$ & $< 0.03$ & $< 0.78$ \\
C$^{++}$ & $>-1.05$ & $>-0.30$ \\
N$^{+}$ & $< 1.08$ & $< 1.83$ \\
N$^{++}$ & $<-0.96$ & $<-0.21$ \\
N$^{+3}$ & $<-0.92$ & $<-0.17$ \\
N$^{+4}$ & $<-0.07$ & $< 0.68$ \\
O$^{+}$ & $< 0.15$ & $< 0.90$ \\
O$^{++}$ & $>-0.75$ & $> 0.00$ \\
O$^{+3}$ & $>-0.76$ & $>-0.01$ \\
O$^{+5}$ & $<-0.54$ & $< 0.20$ \\
Si$^{+}$ & $< 1.34$ & $< 2.09$ \\
Si$^{++}$ & $< 0.12$ & $< 0.87$ \\
Fe$^{++}$ & $< 3.77$ & $< 4.52$ \\
\tableline
\end{tabular}
\end{center}
\end{table}

Because the ionization corrections for \oiii/\ciii\ are nearly
constant at $-0.2$\,dex for the relevant ionization parameter,
we estimate [O/C]~$\lesssim +0.2$ based on the saturated
O\,III~$\lambda 832$ and C\,III~$\lambda 977$ profiles.
In this absorber, at least, it is unlikely the gas has a 
super-solar O/C ratio, in contrast to the abundances inferred
for a sample of $z \gtrsim 1$ absorption systems with
comparable ionization state \citep{vogel95}.
The upper
limit measurements for the nitrogen ions also place a limit on the N/O 
abundance.  In particular, note that the predicted \niii\ column density
closely traces $\N{\oiii}$ for $\log U \approx -2$.  Therefore, the
ratio of these ions provides a measurement of N/O largely
independent of uncertainties in the ionization parameter:  [N/O]~$< -0.2$~dex.
Noting that the $\N{\niii}$ value is likely to be within $\approx 0.1$~dex
of the adopted upper limit (weak absorption is apparent at the expected position
of N\,III~989), we report [N/O]~$= -0.3 \pm 0.15$~dex.  This 
relative abundance follows the [N/O] vs.\ [O/H] trend observed for galaxies
in the local universe \citep[e.g.][]{henry00}.

Because this absorber is characterized by higher ionization states, 
the results are sensitive to the assumed shape of the EUVB radiation
field.  If we adopt the softer model~QG, the observed ionic ratios for
oxygen require a higher ionization parameter ($U \approx -1.4$)
and imply lower elemental abundances ([O/H]~$> -1.2$).  
The latter result is a consequence of the higher H 
ionization fraction of the softer
spectrum.  Again, a measurement of the \civ\ column density would help
resolve this ambiguity.
Additionally, tighter constraints on the column densities of \ovi\ 
and the N ions would probe the shape of the ionizing spectrum.

\begin{figure}
\plotone{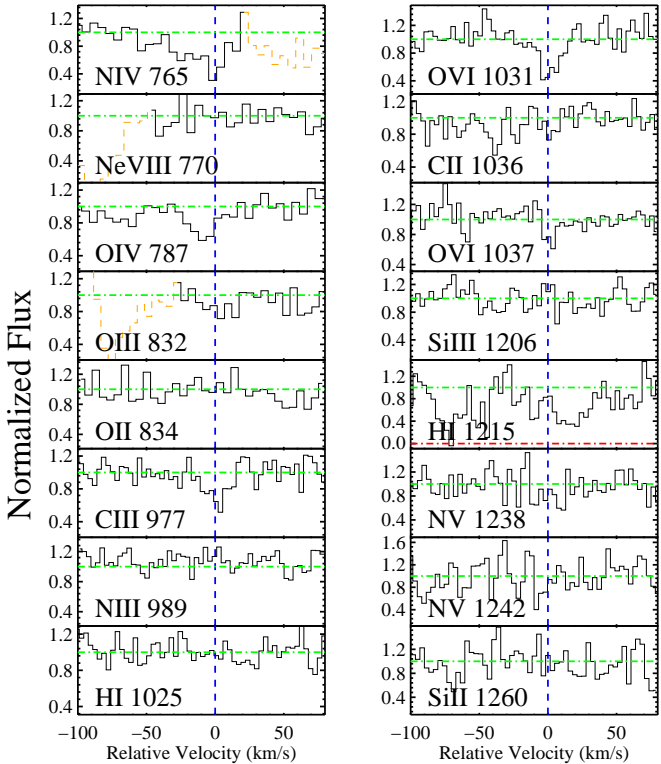}
\caption{
Velocity profiles of the Lyman series and metal-line transitions
analyzed for the absorption system at $z=0.36332$.
}
\label{fig:3633}
\end{figure}

\begin{table}[ht]\footnotesize
\begin{center}
\caption{{\sc IONIC COLUMN DENSITIES FOR THE ABSORBER AT $z$=0.36332\label{tab:z0.3633}}}
\begin{tabular}{lcccc}
\tableline
\tableline
Ion &$\lambda$ (\AA) & EW (m\AA) & AODM & $N_{adopt}$ \\
\tableline
HI & & & & $13.43 \pm 0.10$\\
CII&1036.3367&$<  13 $&$ < 13.22$ &$< 13.22$ \\
CIII& 977.0200&$  22 \pm   7 $&$ 12.64 \pm 0.12 $&$ 12.64 \pm 0.12$ \\
NII&1083.9900&$<  15 $&$ < 13.33$ &$< 13.33$ \\
NIII& 989.7990&$<  19 $&$ < 13.47$ &$< 13.47$ \\
NIV& 765.1480&$  42 \pm   7 $&$ > 13.22$ &$> 13.22$ \\
NV&1238.8210&$<  27 $&$ < 13.34$ &$< 13.35$ \\
NV&1242.8040&$<  31 $&$ < 13.70$ &$$ \\
OII& 834.4655&$<  13 $&$ < 13.33$ &$< 13.33$ \\
OIII& 832.9270&$<  11 $&$ < 13.40$ &$< 13.40$ \\
OIV& 787.7110&$  28 \pm   6 $&$ 13.73 \pm 0.10 $&$ 13.73 \pm 0.10$ \\
OVI&1031.9261&$  23 \pm   8 $&$ 13.44 \pm 0.11 $&$ 13.44 \pm 0.11$ \\
OVI&1037.6167&$<  11 $&$ < 13.45$ &$$ \\
NeVIII& 770.4090&$<  13 $&$ < 13.55$ &$< 13.55$ \\
SiII&1260.4221&$<  31 $&$ < 12.53$ &$< 12.53$ \\
SiIII&1206.5000&$<  24 $&$ < 12.26$ &$< 12.26$ \\
\tableline
\end{tabular}
\end{center}
\end{table}

\subsection{$z=0.36332$}

There is a neighboring absorber (at $+550 \mkms$) to the $z=0.36080$ system
which shows O\,IV, O\,VI, N\,IV, and C\,III absorption yet a weak and 
unusual \lya\ profile (Figure~\ref{fig:3633}).
In particular, one notes that the metal-lines are significantly offset from the
H\,I absorption.  Integrating the \lya\ profile from 
$v = -30$ to $+80 \mkms$, we derive an H\,I column density 
$\N{HI} = 10^{13.4 \pm 0.2} \cm{-2}$.  In the following, we will
adopt this value but one may consider it an upper limit to the H\,I
column density associated with this metal-line system.
It would be extremely valuable to have higher S/N of the C\,IV and
\lya\ spectral regions for this absorber to more confidently
compare their line profiles.  We suggest this system is an example
of an \ovi\ absorber without detected neutral hydrogen gas which
implies a very high temperature and/or metallicity.
Table~\ref{tab:z0.3633} lists the ionic column densities for this absorber
which includes an upper limit on $\N{\niv}$ based on the blended
N\,IV~765 profile.

\begin{table}[ht]\footnotesize
\begin{center}
\caption{{\sc ELEMENTAL ABUNDANCES FOR THE ABSORBER AT $z$=0.36332\label{tab:z0.3633x}}}
\begin{tabular}{lccc}
\tableline
\tableline
Ion & [X/H] & [X/Si$^+$] \\
\tableline
C$^{+}$ & $< 2.36$ & $< 2.34$ \\
C$^{++}$ & $-0.37\pm0.10$ & $-0.39$ \\
N$^{+}$ & $< 3.22$ & $< 3.20$ \\
N$^{++}$ & $< 0.97$ & $< 0.95$ \\
N$^{+3}$ & $> 0.34$ & $> 0.32$ \\
N$^{+4}$ & $< 0.58$ & $< 0.56$ \\
O$^{+}$ & $< 2.59$ & $< 2.57$ \\
O$^{++}$ & $< 0.31$ & $< 0.29$ \\
O$^{+3}$ & $ 0.02\pm0.07$ & $ 0.00$ \\
O$^{+5}$ & $ 0.04\pm0.44$ & $ 0.02$ \\
Ne$^{+7}$ & $< 0.05$ & $< 0.03$ \\
Si$^{+}$ & $< 4.79$ & $< 4.77$ \\
Si$^{++}$ & $< 2.92$ & $< 2.90$ \\
\tableline
\end{tabular}
\end{center}
\end{table}

It is surprising, however, that the absorber also shows C\,III 977.
Regarding collisional ionization equilibrium models, 
there is no single temperature solution
which would give $\N{\ciii}/\N{\ovi} \approx 0.15$ ($T < 2 \sci{5}$\,K)
and \\
$\N{\oiv}/\N{\ovi} \approx 2$ ($T \approx 2.5 \sci{5}$\,K)
assuming solar relative abundances.  Although the difference in these
temperatures may appear small, the relative abundances of 
$\N{\ciii}/\N{\ovi}$ and $\N{\oiv}/\N{\ovi}$ are very sensitive to 
temperature \citep[see also][]{howk04}.
For example, assuming $T = 2.5 \sci{5}$K would
underpredict [C/O] by two orders of magnitude.  Therefore, a single-phase
CIE model cannot fully describe this absorber.
Adopting photoionization as the dominant mechanism, the best
model has $\log U = -1.4 \pm 0.2$ assuming the EUVB model~Q and [M/H]~=0. 
Table~\ref{tab:z0.3633x}
presents the [X/H] and [X/O$^{+3}$] values for $U_{best}=10^{-1.4 \pm 0.2}$.
Even adopting $\N{HI} = 10^{13.4} \cm{-2}$, we calculate a solar
oxygen abundance for this absorber.  Independent of the $\N{HI}$ value
we find [O/C]~$\approx +0.4 \pm 0.2$~dex.
These abundances are comparable to those derived
from the nearby $z=0.36080$ system.  Although the velocity separation
is too large to associate spatially the gas, the `clouds' may
have a common enrichment history suggesting a similar physical origin. 

\vskip 0.5in

\begin{figure}
\plotone{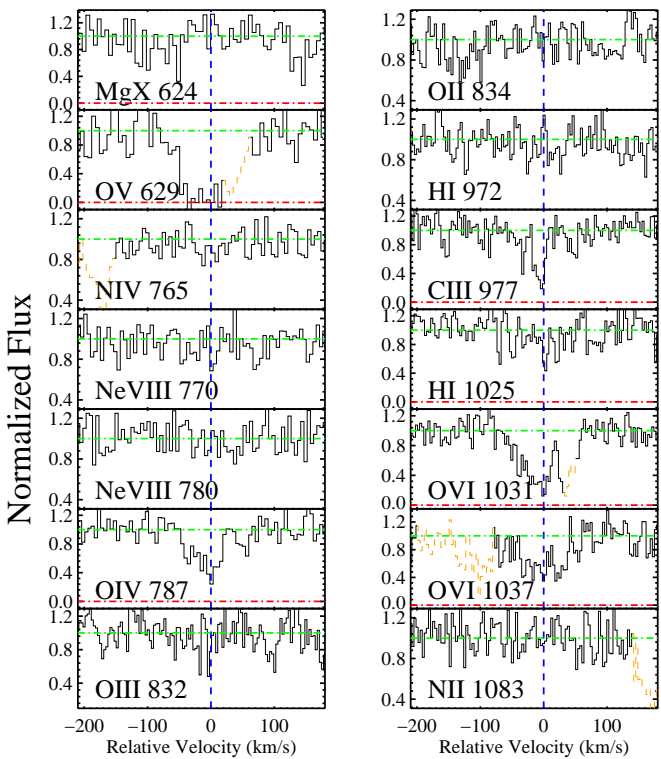}
\caption{
Velocity profiles of the Lyman series and metal-line transitions
analyzed for the absorption system at $z=0.49510$.  
}
\label{fig:4951}
\end{figure}

\begin{table}[ht]\footnotesize
\begin{center}
\caption{{\sc IONIC COLUMN DENSITIES FOR THE ABSORBER AT $z$=0.49510\label{tab:z0.4951}}}
\begin{tabular}{lcccc}
\tableline
\tableline
Ion &$\lambda$ (\AA) & EW (m\AA) & AODM & $N_{adopt}$ \\
\tableline
HI & & & & $14.39 \pm 0.07$\\
CII& 903.9616&$<  20 $&$ < 13.08$ &$< 13.08$ \\
CIII& 977.0200&$  61 \pm   7 $&$ 13.18 \pm 0.06 $&$ 13.18 \pm 0.06$ \\
NII&1083.9900&$<  22 $&$ < 13.47$ &$< 13.47$ \\
NIII& 989.7990&$<  12 $&$ < 13.29$ &$< 13.29$ \\
NIV& 765.1480&$<  16 $&$ < 12.88$ &$< 12.88$ \\
OII& 834.4655&$<  13 $&$ < 13.37$ &$< 13.37$ \\
OIII& 832.9270&$<  15 $&$ < 13.59$ &$< 13.59$ \\
OIV& 787.7110&$ 105 \pm   9 $&$ > 14.34$ &$> 14.34$ \\
OV& 629.7300&$ 164 \pm   9 $&$ > 14.41$ &$> 14.41$ \\
OVI&1031.9261&$ 133 \pm   9 $&$ 14.25 \pm 0.04 $&$ 14.26 \pm 0.03$ \\
OVI&1037.6167&$  94 \pm  10 $&$ 14.31 \pm 0.06 $&$ $ \\
NeVIII& 770.4090&$  26 \pm   7 $&$ 13.75 \pm 0.10 $&$ 13.75 \pm 0.10$ \\
NeVIII& 780.3240&$<  14 $&$ < 13.86$ &$$ \\
MgX& 624.9500&$<  16 $&$ < 14.22$ &$< 14.22$ \\
SVI& 933.3780&$<   9 $&$ < 12.62$ &$< 12.62$ \\
\tableline
\end{tabular}
\end{center}
\end{table}
 
\subsection{$z=0.4951$}

The highest redshift metal-line system along the 
sightline to PKS0405--123 is at $z=0.495$ and exhibits detections
of \ciii, \oiv, \ov, and possibly \neviii\ transitions.
\cite{williger04} performed a profile fit to the \lyb\ transition
and measured $\log \N{HI} = 14.39 \pm 0.07$ and $b=60 \pm 15 \mkms$
consistent with our \EW\ measurements.  
The H\,I and metal-line profiles are presented in 
Figure~\ref{fig:4951} and the ionic column densities are listed in 
Table~\ref{tab:z0.4951}.  

A full analysis of this absorber is presented in \cite{howk04} and we
only summarize a few key results here.  The FUSE spectra provide constraints
on a number of transitions which are very rarely observed in intervening
quasar absorption line systems including O\,V~629, 
N\,IV~765, and the Ne\,VIII and Mg\,X doublets.  Altogether, the FUSE+STIS 
datasets provide 
a comprehensive analysis of the ionization state of this gas.  While a 
collisional ionization model with $T \approx 2.3\sci{5}$~K can reproduce
the observed O ionic column densities, this model underpredicts
$\N{\ciii}$ by over 2 orders of magnitude.  Therefore, 
\cite{howk04} rule out a single-phase CIE model.
In contrast, the gas is well modeled by a single-phase
photoionization model with $\log U \approx -1.2$ although a two-phase
model (collisional plus photoionization) is also allowed.
We summarize the physical properties of this gas in the
following section.

\begin{figure}
\plotone{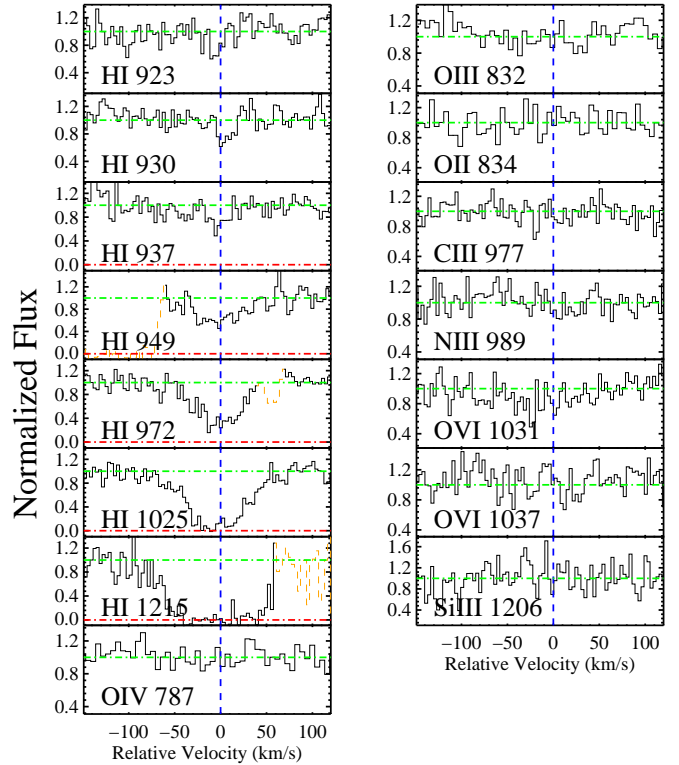}
\caption{
Velocity profiles of the Lyman series and metal-line transitions
analyzed for the absorption system at $z=0.40570$.  Note that 
none of the metal-line transitions show significant detections.
}
\label{fig:4057}
\end{figure}

\section{$\N{HI} > 10^{14} \cm{-2}$ ABSORBERS WITHOUT METAL-LINE TRANSITIONS}
\label{sec:other}

In our analysis of PKS0405--123 we have searched for metal-line
transitions in every \lya\ absorption system identified by
\cite{williger04}, with special attention to $\N{HI} > 10^{14} \cm{-2}$
systems.  This allows us to investigate the metallicity distribution
of a `complete' sample of low redshift, $\N{HI} > 10^{14} \cm{-2}$ \lya\ clouds. 
Perhaps the most remarkable example is the absorber at $z=0.40570$.
Although this absorber shows nearly the same H\,I column density as the
metal-line system at $z=0.36080$, it is remarkable for exhibiting no 
metal-line transitions.  In Figure~\ref{fig:4057} we present the Lyman series
and a set of the strongest metal transitions covered by the FUSE and STIS
spectra (Table~\ref{tab:z0.4057}).  
A COG analysis gives $\N{HI} = 10^{14.85 \pm 0.05} \cm{-2}$
with Doppler parameter $b=34 \pm 2 \mkms$. 
Formally, there is a detection of \ovi\ gas via the broad \ovi~1031 transition,
yet the continuum varies significantly near 1450\AA\ in the STIS
spectrum and the measured value is inconsistent with the upper limit
implied by the \ovi~1037 transition.  In the following, we adopt the
upper limit from the \ovi~1037 transition. 

\begin{table}[ht]\footnotesize
\begin{center}
\caption{{\sc IONIC COLUMN DENSITIES FOR THE ABSORBER AT $z$=0.40570\label{tab:z0.4057}}}
\begin{tabular}{lcccc}
\tableline
\tableline
Ion &$\lambda$ (\AA) & EW (m\AA) & AODM & $N_{adopt}$ \\
\tableline
HI & & & & $14.85 \pm 0.05$\\
CII& 903.9616&$<  13 $&$ < 12.92$ &$< 12.92$ \\
CII&1036.3367&$<  13 $&$ < 13.23$ &$$ \\
CIII& 977.0200&$<  14 $&$ < 12.53$ &$< 12.53$ \\
NII&1083.9900&$<  16 $&$ < 13.32$ &$< 13.32$ \\
NIII& 989.7990&$<  16 $&$ < 13.42$ &$< 13.42$ \\
OII& 834.4655&$<  14 $&$ < 13.37$ &$< 13.37$ \\
OIII& 832.9270&$<  15 $&$ < 13.50$ &$< 13.50$ \\
OIV& 787.7110&$<  12 $&$ < 13.43$ &$< 13.43$ \\
OVI&1031.9261&$<  14 $&$ < 13.63$ &$< 13.63$ \\
OVI&1037.6167&$<  19 $&$ < 13.67$ &$$ \\
SiIII&1206.5000&$<  32 $&$ < 12.34$ &$< 12.34$ \\
\tableline
\end{tabular}
\end{center}
\end{table}

\begin{figure}
\includegraphics[height=3.6in,angle=90]{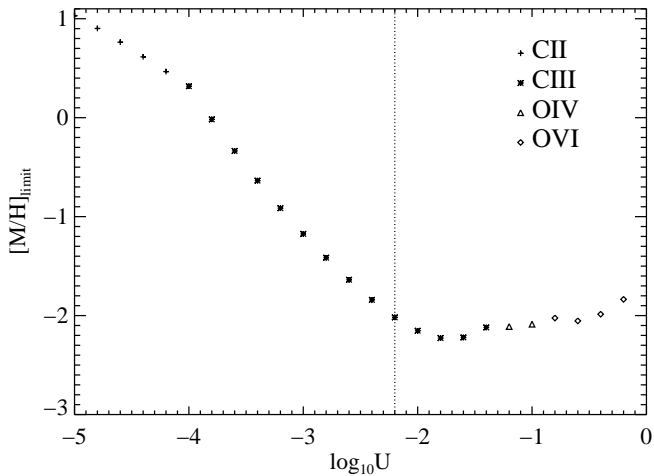}
\caption{Upper limit to the metallicity of the absorber at $z=0.40570$
as a function of ionization parameter.  At each $U$ value,
the point style indicates the
ion with the tightest constraint on the metallicity.  For most of the
parameter space, C\,III places the lowest limit to [M/H].
The dashed line indicates the likely ionization parameter
for an absorber with $\N{HI} = 10^{14.8} \cm{-2}$ ($\S$~\ref{sec:ion}).
}
\label{fig:4057mtl}
\end{figure}

Assuming an equilibrium photoionization model with EUVB model~Q, 
the upper limits to the observed ions constrain the metallicity of the
gas as a function of the ionization parameter.  In Figure~\ref{fig:4057mtl},
we plot the upper limit to the gas metallicity [M/H] for a range of $U$
parameters.  The plot symbols indicate the ion which places the most 
sensitive limit to [M/H].    
For $\log U \approx -3$, the gas must have [M/H]~$<-1$, but this
would imply $n_H \approx 10^{-3} \cm{-3}$ (for $J_{912} = 2\sci{-23}$)
which is considerably higher than expected for such a low $\N{HI}$ absorber.
A more realistic value is $\log U \gtrsim -2$ for 
$\N{HI} = 10^{14.8} \cm{-2}$ ($\S$~\ref{sec:ion}) which implies
[M/H]~$<-2$\,dex.
The results indicate that this absorber has 5 to 100 times lower metal enrichment
than the otherwise similar absorption system at $z=0.36080$.
Unless this absorber is a special example, 
systems with $\N{HI} \approx 10^{15} \cm{-2}$
have large metallicity dispersion at these redshifts. 
We also note that a CIE model would imply even lower metallicities then
the results shown in Figure~\ref{fig:4057mtl}.

As an aside, we stress that 
\ciii\ places the tightest limit on the metal abundance for the
largest range of parameter space.  This ion and its corresponding
C\,III~977 transition deserve greater attention in discussions of the
\lya\ forest.  At higher redshift, the transition is frequently
blended within the forest, yet we emphasize that it is a very powerful
diagnostic of metal enrichment and ionization conditions within absorbers
like the one at $z=0.40570$ \citep[c.f.][]{schaye03}.

In addition to the absorber at $z=0.4057$, there are several
absorbers with no unambiguous, statistically significant metal-line transitions.
These are: 
$z=0.030, \N{HI} = 10^{14.2} \cm{-2}$;
$z=0.351, \N{HI} = 10^{14.2} \cm{-2}$; \\
$z=0.409, \N{HI} = 10^{14.4} \cm{-2}$;
$z=0.538, \N{HI} = 10^{14.2} \cm{-2}$.
In several cases, there are suggestions of metal-line absorption
but line blending or poor S/N prevents a definitive measurement.
In all cases, one can estimate an upper
limit to their metal enrichment with an analysis similar to the
one we performed for the $z=0.40570$ system.  
Under the assumption of photoionization
and assuming $\log U > -2$, we find [M/H]~$<-1.5$ based on
the non-detection of C\,III~977 and/or \ovi\ and adopt this
upper limit in the following discussion.

\begin{table*}
\begin{center}
\caption{{\sc SUMMARY TABLE\label{tab:summ}}}
\begin{tabular}{lcccccccccccc}
\tableline
\tableline
$z$ & $\N{HI}$ & $b_{HI}$ & $\N{OVI}$
& Ion$^a$ & $\log U$ & [M/H]$_{phot}$ & $\log\N{H}$
& $n_H^c$ & $\ell^c$ & $T_{coll}$ & [M/H]$_{coll}$ & [N/$\alpha$] \\
& ($\cm{-2}$) & & $(\cm{-2})$ &&&& ($\cm{-2}$) & ($10^{-5}$) & 
 (kpc) & (10$^5$\,K) \\
\tableline
0.09180 & 14.5 & 38 & 13.8 & ?? & $>-1.5$  & $>-1.4$ & 18.8 & $<2$ & $>90$ & 
     $>2.5$ & $>-2.2$ \\ 
0.09658 & 14.7 & 40 & 13.7 & Photo & $-1.2$  & $-1.5$  & 19.2 & 1.2 & 440 &  
   --           & \\ 
0.16710 & 16.5 & 38 & 14.8 & Multi & $-2.9$  & $-0.25$ & 18.8 & 60 & 3 &  
   --           & & $\gtrsim 0.2$\\ 
0.18250 & 14.9 & 27 & $<13.8$ & ?? &  --      &  --     &  --  &  -- &   $\lesssim -1$   \\
0.18290 & 14.1 & 26$^b$ & 14.0 & Coll & $>-1$   & $>-1$   & $>18.4$ & 3 & $\approx -1.5$ \\
0.36080 & 15.1 & 27 & $<13.3$ & Photo & $-2.0$    & $>-0.7$ & 18.6 & 8 & 16 &  
  --  & & $-0.3$\\
0.36332 & 13.4 & -- & 13.4 & Photo & $-1.4$  &  0      & 17.4 & 2 & 1 &  \\
0.49510 & 14.4 & 60$^b$ & 14.3 & Photo & $-1.3$  & $>-0.3$ & 18.5 & 1.7 & 60 & 
   2.6 & $>-1$ & $<-0.5$ \\
\tableline
\end{tabular}
\end{center}
\tablenotetext{a}{This column lists the dominant ionization mechanism for
the gas as inferred from the metal-line ratios.  In several cases, the
situation is ambiguous or multiple processes are required.}
\tablenotetext{b}{There is a $>25\%$ uncertainty in this value.}
\tablenotetext{c}{These values assume $J_{912} = 2\sci{-23}$\,cgs.}
\end{table*}

\section{DISCUSSION}

Section~\ref{sec:analysis} presented a detailed analysis of the metal-line
systems observed along the sightline to PKS0405--123.  With the combined 
FUSE+STIS datasets, we have obtained constraints on the ionization
mechanism, temperature, metallicity, and elemental abundance ratios
of the gas.  We summarize these properties in Table~\ref{tab:summ}
which lists the redshift, $\N{HI}$ value, $\N{\ovi}$ value, 
the ionization parameter, metallicity [M/H]$_{phot}$, 
the total H column density $\N{H}$ assuming photoionization, 
the mean volume density $n_H$, and an estimate of the length scale $\ell$
assuming photoionization and $J_{912} = 2\sci{-23}$\,cgs. 
The table also lists the temperature and metallicity [M/H]$_{coll}$ for 
collisionally ionized gas, and an estimate of N/$\alpha$,
the abundance of nitrogen relative to an $\alpha$-element (e.g.\ Si, O).
The following sub-sections consider various implications of our
analysis on the nature of metal-line systems in the low redshift
universe.

\subsection{Tracing the WHIM with O\,VI Absorbers}

As noted in the introduction, current theoretical expectation
is that a significant fraction of baryons at $z<1$ are in
a warm-hot intergalactic medium (WHIM). 
One of the few means of detecting this gas is
through \ovi\ absorption.  To this end, Tripp and collaborators
have recently surveyed \ovi\ gas along a number of quasar and AGN sightlines
\citep{tripp00,savage02}.
Their latest results indicate an 
incidence $dN_{\ovi}/dz = 14^{+9}_{-6}$ for an \EW\ limit of 50m\AA.
This implies a baryonic mass density $\Omega_{\ovi} > 0.002 h_{75}^{-1}$ 
assuming the average metallicity is 0.1 solar and a very conservative
ionization correction ($\N{O\,VI}/\N{O} < 0.2$).

With the large redshift of PKS0405--123 and its extensive UV spectroscopy,
we can derive \ovi\ statistics for this sightline.
We have detected six
\ovi\ systems to an \EW\ limit of 30 m\AA\ over a non-contiguous redshift
path of $\Delta z = 0.38$.  This pathlength was determined by identifying
all of the regions free of Galactic H$_2$ and other coincident transitions
with sufficient S/N to identify and measure both members\footnote{ 
Note that these selection criteria exclude the likely \ovi\
system at $z=0.08$ because the \ovi~1037 transition is blended with
a coincident transition.}
of the \ovi\ doublet to a $3\sigma$ \EW\ threshold of 60~m\AA.  
We also searched for any \ovi\ absorbers 
without corresponding \lya\ absorption 
(to the same limit) and found no examples.  
Assuming Poisson statistics \citep[e.g.][]{gehrels86}, 
we find $dN_{\ovi}/dz = 16^{+9}_{-6}$ ($\pm 1 \sigma$)
which is consistent with \cite{savage02}.
If we combine our results with \cite{savage02} -- ignoring the small
difference in sensitivity -- we find: 
$dN_{\ovi}/dz = 15^{+5}_{-4}$ ($\pm 1 \sigma$).

An important aspect of the present work is that our observations
place constraints on the ionization state and/or metallicity of several
\ovi\ absorbers.  For the systems at $z=0.09650, z=0.36332$ and $z=0.49510$,
the detection of C\,III~977 precludes a single-phase CIE model
and therefore suggests the gas is photoionized, in multiple ionization phases,
or in a non-equilibrium state
\citep[see also][]{howk04}.  
If these absorbers have low temperature ($T < 10^5$\,K),
they are unlikely to 
contribute to the WHIM or present a significant reservoir of baryons.
Future observations of the \civ\ profiles of these absorbers would be
particularly valuable to address this issue.
In contrast, the \ovi\ gas in the absorbers at $z=0.16710$ and $z=0.18290$
are well modeled by a single-phase CIE model and we believe this
gas is collisionally ionized.  
Finally, the absorber at $z=0.09180$ is well modeled by both CIE
and photoionization solutions.
If the fraction of photoionized \ovi\ absorbers is roughly half,
then $\Omega_{\ovi}$ in the WHIM is roughly 50$\%$ the
value that one would infer assuming all of the \ovi\ gas is collisionally
ionized.  
It will be important to examine the environment of the absorbers
as a function of ionization state, e.g.\ to determine
if the collisionally ionized gas arises preferentially in galactic
halos \citep{sembach03} or large-scale structures.

Our analysis also constrains the metallicity of the gas giving rise to 
\ovi\ absorption.  For the absorbers presumed to be photoionized, the gas 
metallicity ranges from [M/H]~$\approx -1.5$ to 0\,dex.  The collisionally
ionized gas shows systematically lower metallicity 
([O/H]~$\approx -2$ to $-1$\,dex)
because this gas has much larger hydrogen ionization fraction.  
Altogether, these measurements are consistent with the values assumed
and expected for \ovi\ systems if the gas is related to the WHIM 
\citep[e.g.][]{savage02,dave01b}.  It is worth noting, however, that several
of the \ovi\ lines are at the detection limit of this dataset and, therefore,
absorbers with significantly lower metallicity would not have been detected.
Unfortunately, it may not be possible 
with FUSE or HST/STIS to probe metallicities below [O/H]~$\approx -2$.

\subsection{Tracing the WHIM with Ne\,VIII Absorbers}

At temperatures $T > 5\sci{5}$\,K, \ovi\ is no longer the dominant
ion of oxygen.  This is unfortunate because current simulations of the
WHIM predict the bulk of the gas is at temperatures $T > 10^6$\,K.
For these reasons, X-ray spectroscopy of O\,VII and O\,VIII transitions
offer a more direct probe of the WHIM \citep[e.g.][]{fang02} yet
current technology allows a search for O\,VII and O\,VIII in
only the brightest few X-ray sources.
Even with the next generation of X-ray telescopes, surveys for
O\,VII and O\,VIII will have limited impact.
An alternative means of probing collisionally ionized 
gas with $T \gtrsim 10^6$\,K is
through the \neviii~$\lambda\lambda 770,780$ doublet.  At these
temperatures, \neviii\ is the dominant Ne ion and this relatively strong
pair of transitions can be observed with FUSE and HST/STIS spectroscopy
at $z > 0.2$.  Specifically, FUSE spectroscopy with comparable spectral
quality to the observations presented here will be sensitive to \neviii\ gas 
in absorbers with $\log \N{HI} + {\rm [M/H]} > 13$ for 
$10^{5.5}$\,K~$\lesssim T \lesssim 10^7$\,K.

For sightlines with significant Galactic H$_2$ absorption
(PKS0405--123 is a moderate example), a survey for \neviii\ is difficult
for $\lambda < 1000$\AA\ or $z < 0.3$.  Nevertheless, we have conducted
a search for \neviii\ absorption by examining the Ne\,VIII regions
for all \lya\ absorbers detected along the PKS0405--123 sightline.  
We found no significant detections.
Furthermore, for $\lambda > 1000$\AA\ we have found no pairs of
absorption features with the appropriate separation of the \neviii\
doublet, independent of the presence of a corresponding \lya\ feature.  
Altogether, we estimate the non-contiguous pathlength 
searched to be $\Delta z_{\neviii} = 0.11$.  Assuming Poisson statistics,
we set a conservative $95\%$ upper limit to the
incidence of \neviii\ absorption: $dN_{\neviii}/dz < 40$ 
for a $3\sigma$ rest-EW limit of $\approx 40$m\AA.
This result will be extended to lower \EW\ limits and pursued along additional
sightlines with forthcoming, approved FUSE programs (PI: Howk). 
We will also extend the analysis
to include a search for Mg\,X absorption for systems with
$z > 0.5$.

\subsection{Ionization of the IGM:  A Possible $\N{H}$ Conspiracy}
\label{sec:ion}

The analysis described in $\S$~\ref{sec:analysis} allows us to
investigate global trends in the ionization state of low $z$ metal-line
systems.   One key result is that a significant fraction of the
systems are predominantly photoionized.  This is not a surprising
result; current expectation is that the majority of the low $z$ \lya\
forest is photoionized gas \citep[e.g.][]{dave01}.  
Unlike studies of the majority of \lya\ `clouds', however, 
an analysis of the metal-line systems provides 
an assessment of the ionization fraction and metallicity
of this gas.

\begin{figure}
\begin{center}
\includegraphics[height=3.6in,angle=90]{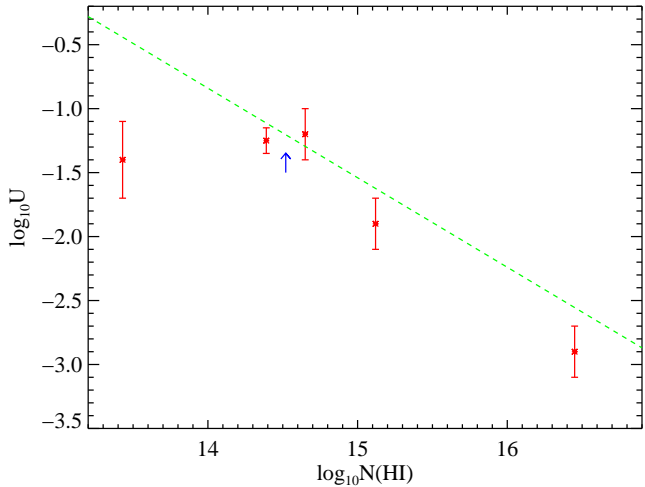}
\caption{$\log U$ vs. $\log\N{HI}$ for the systems expected to
be predominantly photoionized.  The dashed line indicates a theoretical
prediction based on comparisons between
CDM numerical simulations and the low redshift \lya\ forest \citep{dave01}.
}
\label{fig:UNHI}
\end{center}
\end{figure}

Focusing on the 
subset of systems where observations constrain the ionization state, 
we can examine trends between $U$ and $\N{HI}$.
Figure~\ref{fig:UNHI} presents a plot of $\log U$ vs $\log \N{HI}$ for
all of the systems where we believe photoionization is the dominant mechanism.
We caution, again, that in several cases the presumption of photoionization
is not secure.  In particular, the gas in the absorbers toward $z=0.36332$ 
and $z=0.49510$ may be in multiple ionization phases.
In one case, we have only placed a lower limit to the $U$ value; 
this is designated by the up arrow.  Examining the figure, 
one notes a general decline in $U$ with increasing $\N{HI}$.
Performing the conservative generalized Spearman test, 
we calculate the significance to be $85\%$. 
Similarly, a Pearson test adopting the lower limit as a value rules out the null
hypothesis at $97\%$~c.l. and the results are more significant if
we exclude the `anomalous' absorber at $z=0.36332$.

Overplotted in the figure (dashed-line) 
is a prediction for the correlation between
$U$ and $\N{HI}$ at $z=0.2$ derived in the following way.
First, we adopt the 
\cite{dave99} relation between volume density $n_H$ and $\N{HI}$
for low $z$ \lya\ absorbers:  

\begin{equation}
1+\delta \equiv \frac{\rho}{\bar\rho} = \frac{n_H}{{\bar n}_H} 
\propto \N{HI}^{0.7} 10^{-0.4 z} 
\label{eqn:corr}
\end{equation}

\noindent where

\begin{equation}
{\bar n}_H \equiv \mu \Omega_b \rho_c (1+z)^3
\label{eqn:nH}
\end{equation}

\noindent is the mean hydrogen density of the universe.  Second,
we assume a constant H\,I ionizing flux.  This gives 
$U \propto \N{HI}^{-0.7}$ and sets the slope of the dashed 
line in Figure~\ref{fig:UNHI}.  Finally, the normalization of the line is 
determined by assuming a value for the H\,I photoionization rate;
the central value reported by \cite{dave01} from their analysis
of the statistical properties of the $z < 1$ \lya\ forest 
($\Gamma_{HI} = 10^{-13.3}$\,photons~$\rm s^{-1}$ for $\bar z = 0.17$)
gives the result shown here.

The correspondence between observation and theory is impressive.
The curve is not a fit
to the observations, it is a a comparison of prediction
from numerical simulations with observed properties 
of the low $z$ \lya\ forest.
It is especially noteworthy that 
the curve results from an analysis of the low $z$ \lya\ forest 
while our results
are derived from the metal-line systems presented in this paper.
Of course, the metal-line systems are simply a subset of the \lya\
forest although the metal-line sample
exhibits $\N{HI}$ values which exceed
the principal range of the \cite{dave99} analysis.
Furthermore, we caution that the majority of metal-line systems have
$\N{HI} \approx 10^{14.5}$, i.e.\ the extremes of the parameter space
described by Figure~\ref{fig:UNHI} are sparsely sampled.
Nevertheless, the results presented in Figure~\ref{fig:UNHI}
support the model of the IGM inferred from cosmological simulations.

\begin{figure}
\includegraphics[height=3.6in,angle=90]{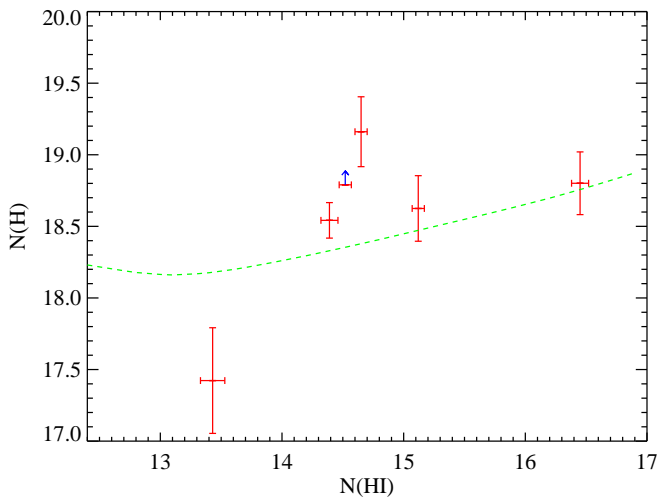}
\caption{Inferred total hydrogen column densities $\N{H}$ as a function
of $\N{HI}$ value for the metal-line systems toward PKS0405--123 which
we believe are predominantly photoionized.  The dashed line in 
Figure~\ref{fig:UNHI} maps to the dashed line shown here.
Remarkably, both observation and theory suggest the \lya\ forest
clouds have roughly uniform $\N{H}$ value for $\N{HI} = 10^{12-16} \cm{-2}$.
}
\label{fig:NH}
\end{figure}

Pushing the analysis one step further, we find a startling result
regarding the total hydrogen column density $\N{H}$ of the low $z$
\lya\ forest.
Examining Figure~\ref{fig:NH} and 
column 6 of Table~\ref{tab:summ}, one notes that the
predicted $\N{H}$ values for the metal-line systems based on 
photoionization modeling are nearly independent of $\N{HI}$.
For all but one of the absorbers\footnote{The one exception 
is the unusual \ovi\ absorber
at $z=0.3633$ which shows a very weak \lya\ profile apparently
offset from the metal-line transitions.  We include it here, but
note its peculiar nature.}, $\log \N{H} = 18.75 \pm 0.3$\,dex.
While this result could be biased by our reliance on 
metal-line systems (e.g.\ lower $\N{H}$ systems
may not be detected unless the gas
had a very low ionization fraction),  the roughly
constant $\N{H}$ value is a natural consequence of 
the correlation between $n_H$ and $\N{HI}$. 
Higher $\N{HI}$ implies higher $n_H$ which
implies lower $U$ which implies a lower ionization fraction and, finally,
a lower $\N{H}/\N{HI}$ ratio.   In this manner, the correlation between
$n_H$ and $\N{HI}$ can lead to a roughly constant $\N{H}$ value.

The dashed line in Figure~\ref{fig:NH} presents the predicted $\N{H}$
values as a function of $\N{HI}$ assuming (1) the EUVB model~Q at $z=0.3$
and (2) $\log U = -3 - [\N{HI}/10^{16.5}]^\beta$ with $\beta = 0.7$.
The latter is equivalent to assuming a constant intensity for the EUVB
radiation field and adopting the $n_H \propto \N{HI}^{0.7}$ relation
from \cite{dave99}.
We reach the unsettling conclusion
that {\it all \lya\ `clouds' have roughly the same $\N{H}$ value.}
We stress that this result is sensitive to the value of $\beta$;
larger ($\beta > 1$) or smaller $(\beta < 0.5)$ values imply
steeper relations between $\N{H}$ and $\N{HI}$.
One seeks a physical interpretation -- if it exists -- for identifying
$\N{H} \approx 10^{18.5-19} \cm{-2}$
as a fundamental scale for the low redshift \lya\ forest.
This $\N{H}$ value can be identified with a comoving length scale by adopting
the mean hydrogen density of the universe at $z=0$.  Taking 
$\Omega_b = 0.04 h_{70}^{-2}$ \citep{omeara01}, $\N{H} = 10^{19} \cm{-2}$
corresponds to $\ell_{\rm Ly\alpha} \approx 17$~Mpc at $z=0$.  
The observed variations in $\N{HI}$, therefore, result from
the compression of gas along this pathlength in overdense regions.

It is worth noting that the description of the IGM developed
by \cite{schaye01} based on Jeans length arguments also predicts
a shallow dependence of $\N{H}$ with $n_H$ and $\N{HI}$. 
In a future paper, we will investigate this $\N{H}$ conspiracy
through an analysis of numerical simulations and analytic arguments.
If the near constancy of $\N{H}$ is confirmed observationally,
it could challenge models of the \lya\ forest at $\N{HI} < 10^{16} \cm{-2}$ 
which associate the `clouds' with individual galaxies 
\citep[e.g.\ galactic halos;][]{lzt95,chen98,chen01,manning03}.

In passing, we also note that because \cite{dave99} find 
$n_H \propto \N{HI}^{0.7}$ to at least $z=3$, one predicts
the individual \lya\ forest `clouds' have a characteristic 
$\N{H}$ at any given redshift from $z = 0 - 3$!
The characteristic value of $\N{H}$ does vary with redshift owing
to evolution in the EUVB intensity, 
the redshift dependence of Equation~\ref{eqn:corr}, and the increasing
density of the universe (Equation~\ref{eqn:nH}).  
In general, $\N{H}$ is predicted to increase from $z=0$ to $z=3$.


Before concluding this sub-section, we note that we 
searched the literature for additional photoionized, 
metal-line systems to test the above conclusions.
Unfortunately, this resulted in only two cases. 
One of these -- the absorber at $z=0.068$ toward PG0953+415 
\citep[][; $\N{HI}=10^{14.35}\cm{-2}$, $U=-1.35$]{savage02} 
-- has $\N{HI}$ and $U$ values which place it along the correlation
expressed by the metal-line systems toward PKS0405--123 
(Figure~\ref{fig:UNHI}).  Although the
authors did not report an $\N{H}$ value, adopting their ionization
parameter we estimate $\N{H} = 10^{18.5} \cm{-2}$.
near $10^{18.5} \cm{-2}$.  The only other example is the absorber
at $z=0.0053$ toward 3C~273 which lies within the Virgo supercluster
\citep{tripp02}.  Taking their single-phase solution, this absorber
falls below the relation in Figure~\ref{fig:UNHI}, although 
its unique surroundings may imply an unusual physical environment
\citep{stocke04}.
It may also be noteworthy that the two outliers from the $U$, $\N{HI}$
relation fall beneath the curve.  This is consistent with the
results of \cite{dave99} who emphasized the simulations show a 
significant scatter in the $1+\delta \propto \N{HI}^{0.7}$ correlation
at low $z$ with the majority of `outliers' identified above the relation
(i.e.\ lower $U$).

\subsection{Chemical Enrichment of the Low $z$ IGM}
\label{sec:metal}

An active area of current research is the study of the chemical
enrichment of the IGM \citep[e.g.][]{songaila01,pettini03,schaye03}.
These studies test scenarios of Population III star formation
\citep[e.g.][]{gnedin97,wass00} as well as feedback processes in early
galaxy formation \citep{aguirre01}.
Preliminary studies with 10m-class telescopes argued for significant
enrichment in \lya\ `clouds' with $\N{HI} > 10^{14} \cm{-2}$ \citep{tytler95}
but a stacked spectrum of lower H\,I systems showed no corresponding
\civ\ absorption
\citep{lu98}.  Recently, analyses of the `pixel technique'
have argued for metal enrichment to very low overdensity
\citep[e.g.][]{ellison00}.
Recently, \cite{schaye03}
have argued that the \lya\ forest has a median carbon metallicity
[C/H]$~= -3.47^{+0.07}_{-0.06} + 0.08^{+0.09}_{-0.10} (z-3) + 
0.65^{+0.10}_{-0.14} (\log\delta-0.5)$ where $\delta$ is the overdensity
of the gas and is related to $\N{HI}$ as described in 
Equations~\ref{eqn:corr},\ref{eqn:nH}.
\cite{simcoe04} have reached similar conclusions, although 
with $\approx 0.3$\,dex higher values for [O/H] and a less dramatic
dependence on overdensity\footnote{These differences are primarily
due to the fact that the authors' fiducial models assumed EUVB
radiation fields with differing spectral shapes.}.
Very little research, in comparison, has been pursued at low 
redshift.  \cite{burles96} surveyed \ovi\ absorbers at $z \sim 1$
and argue that the metallicity in these systems is [O/H]~$\geq -2.4$\,dex.
Similarly, \cite{barlow98} examined a large sample of \civ\ absorbers
and found a mean metallicity [C/H]~$\approx -1.5$\,dex.  These results,
imply higher enrichment levels within the \lya\ forest at low $z$
and therefore a significant evolution in the chemical enrichment of the IGM.

The analysis presented in this paper (summarized in Table~\ref{tab:summ})
can be used to evaluate the chemical abundances of the low $z$
\lya\ forest for comparison with high $z$.  
We will focus here on absorbers with $\log \N{HI} > 14$
which includes the absorbers without significant metal-line
transitions ($\S$~\ref{sec:other}).
For the systems with detectable metal-line
absorption, the median metallicity lies at $\approx 1/10$ solar enrichment
if the absorbers are predominantly photoionized.
A significant fraction of these absorbers 
exhibit chemical enrichment
comparable to the Milky Way and its Magellanic Clouds.  Another
significant result is the large scatter in chemical enrichment among
the absorbers.  The systems span at least two decades in metallicity
ranging from approximately solar metallicity to $< 1/100$ solar.
Although UV spectroscopy has poorer resolution and S/N than 
optical spectra, we emphasize that the relatively
high metallicities derived here are not the result of observational bias.
For most of the systems, these transitions are saturated or
detected well above threshold.  If this gas had 10$\times$ lower metallicity,
several transitions would still be detectable, although constraints
on the ionization state would be weaker.
If we include the $\approx 5$ absorbers without significant metal-line 
absorption and assume their enrichment is below $-1.5$\,dex, then
the median metallicity lies closer to 1/30 solar.
In either case, the median metallicity
is consistent with the results presented by
\cite{barlow98} based on their analysis of C\,IV absorption in the
$z \sim 0.5$ \lya\ forest.  

Altogether, the results from the PKS0405--123 sightline
indicate that \lya\ clouds with 
$\N{HI} > 10^{14} \cm{-2}$ show a median value of $1/30 - 1/10$ solar
and a log normal dispersion of $\approx 1$\,dex.  This median metallicity
is significantly higher than the value ascribed to \lya\ `clouds' with
comparable $\N{HI}$ values at high $z$ \citep[e.g][]{songaila01,schaye03}. 
In fact, the value and scatter are
remarkably similar to values derived for $z \approx 2$
damped \lya\ systems \citep[e.g.][]{pro03}.
Within the CDM paradigm, however, the `clouds' in the low redshift
\lya\ forest have much higher overdensity $\delta$ compared to high
$z$ \lya\ clouds.  
If a correlation between metallicity and overdensity exists, 
we must be careful to compare metallicities at the same overdensity.
We can derive the metallicity of absorbers with comparable overdensity
in the $z \sim 3$ universe by (i) calculating $1+\delta$ for
an absorber with $\N{HI}=10^{14.5} \cm{-2}$ using Equation~\ref{eqn:corr}
and (ii) evaluating the metallicity 
[C/H] at this overdensity using the results presented by \cite{schaye03}.
We find [C/H]~$= -2.9$ at $z=3$ with a log normal scatter of $0.55$\,dex.
This metallicity and scatter are lower than the sample of absorbers
analyzed in this paper.  Even if we adopt non-solar relative abundances
\citep[e.g. $\lbrack$Si/C$\rbrack$~$= +0.7$;][]{aguirre03},  
the metallicity of these
low $z$ \lya\ absorbers greatly exceeds absorbers with comparable
overdensity at high redshift.

In short, our observations suggest a significant enrichment of
the $\delta \approx 20$ IGM between $z=3$ and the present universe.
This is in stark contrast to current claims that the IGM has at
most a moderate metal enrichment from $z=2$ to 5 
\citep{songaila01,pettini03,schaye03}.
Of course, our range of redshifts corresponds to $\approx 10$\,Gyr 
wherein significant galactic enrichment has occurred.
It apparently requires that the overdense IGM is polluted by processes related
to galaxy formation, e.g., 
winds driven by supernovae and/or mergers \citep[e.g.][]{cox04}.
By inference, this weakens arguments that the IGM has been 
enriched primarily by an initial generation of Population III stars
\citep[e.g.][]{gnedin97,wass00}, although it does not preclude the 
occurrence of both early and ongoing enrichment.

Given a correlation exists between $\N{HI}$ and $n_H$,
one might expect 
correlations between $\N{HI}$ and metallicity.
For example, regions of higher overdensity will
on average be closer to galaxies and therefore may be at higher
metallicity than less dense gas.  As noted above, \cite{schaye03} report
a correlation between metallicity and overdensity when matching
their simulations against C\,IV absorption in the high $z$ \lya\ forest.
In the sample of $z<0.5$ metal-line systems toward PKS0405--123,
however, there is no obvious correlation between $\N{HI}$ and
metallicity. While the partial
LLS at $z=0.16710$ does show near solar abundance, the lower $\N{HI}$
absorbers at $z=0.49510$ and $0.36080$ have comparable metallicity.
Furthermore, there is no apparent correlation of metallicity with
ionization parameter or redshift.
Because these conclusions are based on a small sample, an underlying
trend may be masked by only a few outliers or systematic effects.
A goal of future studies will be to obtain complete observations
for a sample of $N>20$ absorbers with a wide range of $\N{HI}$ values.

In a smaller sample of the PKS0405--123 metal systems, we were able to
constrain the N/O or N/Si relative abundance.  Table~\ref{tab:summ} summarizes
these results. The
majority of absorbers are consistent with the general pattern of N/O values
expressed by local measurements of H\,II regions and stars \citep{henry00}.  
The obvious exception is the absorber
at $z=0.49510$ which shows a super-solar metallicity and an upper limit
to N/O of $-0.6$\,dex under the assumption of photoionization.  
The low N/O value is characteristic of the values
measured in starbursting galaxies \citep[e.g.][]{contini02}
and suggests this gas was enriched by an outflow from such a system. 
Future observations of N\,V and a deeper search for galaxies at
$z=0.495$ will help clarify this picture.

\section{SUMMARY}

We have presented an analysis of nine metal-line absorption systems
toward PKS0405--123 using the combination of {\it FUSE} and HST/STIS
ultraviolet spectroscopy.  We derive H\,I column densities
$\N{HI}$ from curve-of-growth analyses and examine ionic ratios
to constrain the ionization mechanism of the gas.
These ratios are compared against single-phase model predictions
for collisional ionization and photoionization to 
constrain the temperature and/or ionization parameter $U$
and thereby infer physical conditions of the absorber (density,
metallicity, ionization fraction).

We identify six \ovi\ absorbers to a $3\sigma$ equivalent width limit
of $30$m\AA\
along a non-contiguous redshift path $\Delta z_{\ovi} = 0.38$.
This gives $dN/dz|_{\ovi} = 16 \pm 6$, consistent with other low
$z$ sightlines.  These systems exhibit metallicities $\lbrack$O/H$\rbrack$
from $-1.5$ to 0\,dex which roughly matches prediction for the
warm-hot intergalactic medium (WHIM).  Half of the \ovi\ absorbers, however,
show transitions which suggest the gas is photoionized or has multiple
ionization phases.  This may argue against
their membership in the WHIM, although a quantitative assessment awaits
a careful assessment of the ionization state of the gas in the numerical
simulations. We also survey the WHIM through a search for \neviii\ absorbers
which should probe gas at temperatures $T \approx 10^6$\,K.  Over
a non-contiguous pathlength $\Delta z_{\neviii} = 0.11$ we find no such
systems and estimate $dN/dz|_\neviii < 40$ to an $3\sigma$
equivalent width limit of $\approx 40$m\AA.

The photoionized metal-line systems show decreasing ionization parameter
with increasing $\N{HI}$ value.  Both the slope and normalization of this
correlation match a prediction derived from previous analysis of
the \lya\ forest with cosmological simulations \citep{dave01}.  
This correspondence
lends support to paradigm of the IGM described by numerical simulations.  
Furthermore, the observed and predicted $U$ vs.\ $\N{HI}$ correlation
imply a startling `$\N{H}$ conspiracy':  the entire population of \lya\
clouds have roughly identical total hydrogen column density, i.e.\
independent of $\N{HI}$.  We argue this conspiracy holds at all redshifts
although with the characteristic $\N{H}$ value increasing with redshift.

The median metallicity of absorbers with $\N{HI} > 10^{14} \cm{-2}$
is $\approx 1/30 - 1/10$ solar metallicity with a large ($\approx 1$\,dex)
dispersion.  This metallicity is consistent with a previous analysis
of the low $z$ \lya\ forest and indicates a significant enrichment of the
IGM since $z \sim 2$.  There is no obvious correlation between metallicity
and any other physical property of the gas.  Finally, we present a 
small sample of N/O measurements which are generally consistent with
H\,II regions and stars of comparable metallicity.

\acknowledgments

This work is based on observations obtained by the NASA-CNES-ESA {\it FUSE} 
mission, operated by Johns Hopkins University.  It is also based
on observations with the NASA/ESA {\it Hubble Space Telescope}
obtained at the Space Telescope Science Institute, which is 
operated by the Association of Universities for Research in Astronomy, Inc.,
under NASA contract NAS5-26555.
The authors wish to thank P. Madau, A. Aguirre, and R. Dav\'e
for helpful discussions.
This work was supported by a FUSE GI grant to JXP, HWC, and BJW under
NASA contract NAG5-12743.  HWC also acknowledges support by NASA
through a Hubble Fellowship grant HF-01147.01A from the Space Telescope
Science Institute, which is operated by the Association of Universities
for Research in Astronomy, Incorporated, under NASA contract
NAS5-26555. Finally, JCH acknowledges support to NASA contract NAG5-12345.

\clearpage


%
%
%
%
%



\clearpage




\clearpage





\clearpage


\clearpage

\clearpage
 
\clearpage

\clearpage

\clearpage

\clearpage

\clearpage

\end{document}